# Classical and Bayesian statistical methods for low-level metrology

# Classical and Bayesian statistical methods for low-level metrology

MANIFICAT Guillaume

HELALI Salima

BOUISSET Patrick

# Table of contents

# Illustrations

*"It ain't what you don't know that gets you into trouble. It's what you know for sure that just ain't so."*

**Mark Twain**

*"The greatest enemy of knowledge is not ignorance, it is the illusion of knowledge."*

**Stephen Hawking**

# Abstract

This document presents the statistical methods used to process low-level measurements in the presence of noise. These methods can be classical or Bayesian. The question is placed in the general framework of the problem of nuisance parameters, one of the canonical problems of statistical inference. By using a simple criterion proposed by Bolstad (2007), it is possible to define statistically significant results during a measurement process (act of measuring in the vocabulary of metrology). This result is similar for a classic paradigm (called "frequentist") or Bayesian: the presence of zero in the interval considered (confidence or credibility). It is shown that in the case of homoskedastic Gaussians, the commonly used results are found. The case of Poisson distributions is then considered. In the case of heteroscedastic Gaussians, which is that of radioactivity measurement, we can consider them as Poisson laws in the limit of large counts. The results are different from those commonly used, and in particular those from standards (ISO 11929). Their statistical performances, characterized by simulation, are better and are well verified experimentally. This is confirmed theoretically by the use of the Neyman-Pearson lemma which makes it possible to formally determine the statistical tests with the best performances. These results also make it possible to understand the paradox of the possible divergence of the detection limit. It is also formally shown that the confidence intervals thus calculated by getting rid of the nuisance parameter according to established methods result in the commonly used confidence interval. To our knowledge, this constitutes the first formal derivation of these confidence intervals.

This method is based on keeping the measurement results whether they are significant or not (not censoring them). This is recommended in several standards or documents, is compatible with the ISO 11929 standard and is in line with recent proposals in the field of statistics. On the other hand, all the information necessary to determine whether a measurement result is significant or not remains available. The conservation and restitution of all results is currently applied in the USA. The textbook case of the WIPP incident makes it possible to ensure favorable public perception.

The implications and applications of this method in different fields are finally discussed.

# 1. INTRODUCTION

The use of characteristic limits in radiation metrology (decision thresholds and detection limits) commonly leads to the consideration that results below these limits are unusable or meaningless. The situation considered is that of two measurements (measurements in metrology jargon):

• The first is that of a reference measurement in the absence of the desired signal.

• The second is that of a sample with the potential presence of a signal.

From the first reference measurement, characteristic limits are determined below which the signal is assumed to be absent. Below these limits, the measurement result is almost unused or to give an upper limit to the signal. In the last chapter of his book (Willink, 2013), Willink addresses measurement near a limit (noisy low-level signal for example) under the title "Measurement near a limit – an insoluble problem? ". He lists the difficulties encountered and is very pessimistic about the possibility of resolving the numerous paradoxes, inconsistencies and difficulties of this problem whatever the statistical paradigm used (Bayesian or frequentist).

Yet, in other domains, exploitation of data below characteristic limits is universally adopted (James & Roos, 1991).

this document presents the work carried out on the exploitation of low-level measurements in metrology, using the classical paradigm (known as frequentist) and the Bayesian paradigm.

This problem is placed in the more general framework of the elimination of nuisance parameters (Cox & Hinkley, 1974; Liseo, 2005) where the characteristic parameter of the noise (the reference) is not known precisely and is not intrinsically of interest. In fact, only the signal interests us.

After a presentation of the problem framework for each paradigm, we define the characteristic limits (decision threshold and detection limit) in each case. These notions are then applied in the case of homoscedastic Gaussians. The case of Poisson distributions is then presented. At the limit of large counts, these Poisson laws become heteroscedastic Gaussians. The method proposed here is based on the determination of confidence or credibility intervals using conditional and marginal likelihood, which makes it possible to eliminate the nuisance parameter. The presence of the zero value is sufficient to make the measurement non-significant, implicitly defining the characteristic limits. Providing the measurement result and its uncertainty is therefore necessary and sufficient. We will examine the compatibility with current standards and the impact of this method on them. Of course, an exact and numerical evaluation will be made of the statistical performance of this method. It is interesting to place this proposal within the framework of discussions on the concept of statistically significant result which animates the scientific world. The implications will then be examined before concluding.

# 2. PRESENTATION OF THE PROBLEM

We are in the presence of a sample whose measurements are represented by a random variable $G$ associated with a parameter (the measurand which is denoted with a Greek letter μ).
We seek to determine the presence of a signal from one or more measurements carried out on this sample. This presence or absence of signal is determined in relation to a reference $B$ also measured. This reference can be a sample, a measuring installation, etc. It is supposed not to contain a signal. In fact, everything is done physically to be as certain as possible to have a reference where any signal is assumed to be absent.
We seek to determine the difference between the sample and the reference by excess of the measurand compared to the reference. An example of this situation could be the measurement of an activity in a sample against a supposed reference (or not) devoid of any activity. The measurand of the possible signal $S$ is assumed to be able to physically only take strictly positive values (for example mass or activity).

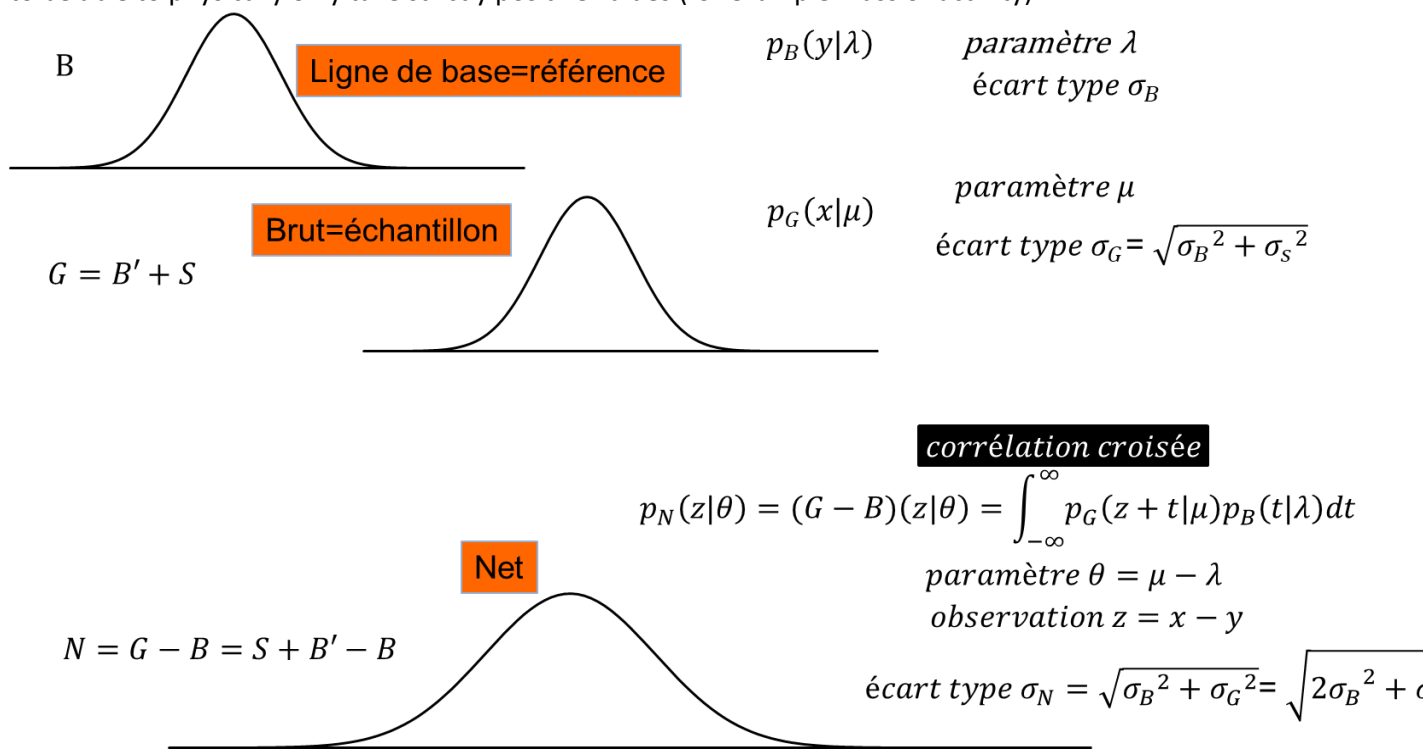

Figure 1 - Frequentist principle diagram of the problem

From the perspective of random variables, the sample is the result of the sum of the reference and the signal we are looking for:

$$G = B' + S$$

Note that in the random variable $G$, we do not know the contribution of $B'$.
In particular, this random variable has two components in its uncertainty: the measurement uncertainty and the possible intrinsic uncertainty of the desired signal. Thus, in the case of radioactivity measurement, the radioactive decay process is intrinsically random and has its own uncertainty independent of the measurement process.

Let us now consider the experimental context: by measuring the reference $B$, we obtain a value $y$ while the measurement of the sample will give a value $x$, realization of the random variable G.

Metrology focuses on the net value $z = x - y$, by definition and with the notations above, it is a realization of the random variable $N = G - B = S + B' - B$.

This random variable $N$ should not be confused with the random variable S even if $B - B'$ is a random variable with zero expectation by definition. B and B' are in fact two independent and identically distributed random variables (iid in the jargon). Remembering that random variables S, B et B' are by definition independent (signal independent of noise), we have in terms of expectation:

$$E(N) = E(G - B) = E(S + B' - B) = E(S) + E(B) - E(B') = E(S)$$

*(1)*

But in terms of variances:

$$Var(N) = Var(G - B) = Var(S + B' - B) = Var(S) + Var(B) + Var(B') = Var(S) + 2.Var(B)$$

*(2)*

By measuring $z$, we therefore have access to the inference on the distribution of the random variable N which will have a greater uncertainty than that of S. By the convolution of S and $B' - B$, we obtain a more dispersed ("spread") random variable.

In particular, if the random variable S is necessarily positive (of positive support more precisely), the random variable N has no reason to be positive.

Thus, we can take for granted the fact that a measurement $z$, realization of the random variable N, can give a negative result (compared to the reference) simply because a subtraction is carried out which, due to intrinsic fluctuations, can lead to a negative value. This point is universally accepted by statisticians and measurement theorists (but not by all metrology practitioners)(Chambless et al., 1992; Ellison, 2014; ISO, 2010a; James & Roos, 1991) (IUPAC, 1998) (EURACHEM, 2012) (CETAMA, 2014) (ISO 11843, 2000).

The following chapter will focus on the statistical characterization of the random variable $N$ in the particular case of a Gaussian distribution of the signal and the noise, to the inference of its parameters as well as to the properties of associated quantities in particular the decision threshold and the detection limit.

# 3. CLASSICAL FREQUENTIST APPROACH IN THE CASE OF HOMOSCEDASTIC GAUSSIAN

let's remember that $N = S + B' - B$ and suppose that the random variables S, B and B' are all gaussians.

## 3.1. Statistical distribution of the variable N in the case of a signal and a noise that are both Gaussian

$$p_S(x|\theta, \sigma_S) = \frac{e^{-\frac{(x-\theta)^2}{2\sigma_S^2}}}{\sqrt{2\pi}\sigma_S}, \quad p_B(x|\lambda) = \frac{e^{-\frac{(x-\lambda)^2}{2\sigma_B^2}}}{\sqrt{2\pi}\sigma_B} \text{ et } p_{B'}(x|\lambda) = \frac{e^{-\frac{(x-\lambda)^2}{2\sigma_B^2}}}{\sqrt{2\pi}\sigma_B}.$$

A linear combination of Gaussian variables being itself Gaussian, we deduce that (Bromiley, 2003) :

$$p_N(z|\theta) = \frac{e^{-\frac{(z-\theta)^2}{2(\sigma_S^2 + 2\sigma_B^2)}}}{\sqrt{2\pi(\sigma_S^2 + 2\sigma_B^2)}}$$

From this distribution, we can deduce two quantities of interest, important in metrology: the decision threshold and the detection limit.

## 3.2. Decision thresholds

### 3.2.1. Definition

In the frequentist paradigm, the decision threshold is the fixed value of the measurand such that, when the measurement result of a measurand quantifying the physical phenomenon is greater than it, we decide that the physical phenomenon is present (ISO, 2010a). Below this value, the measured value could therefore be reasonably explained by a simple fluctuation in the background noise. This threshold is generally determined by hypothesis tests using the Neyman-Pearson methodology (Neyman & Pearson, 1933) formalized in the case of metrology by Currie (Currie, 1968, 1999b, 2000, 2004).

This methodology is based on purely frequentist concepts (fixed parameter). We will therefore reason in the space of observations.

We want to know up to what measured value can we consider that the hypothesis of a zero parameter (no physical phenomenon due to the signal) is reasonable. This is the definition of the critical value of an hypothesis test (Lehmann & Romano, 2005a)

### 3.2.2. Currie approach

In this approach, we identify a hypothesis that we want to test, called the null hypothesis (which we will designate by $H_0$).

It is generally an assertion about a distribution that we wish to test in the form of the absence of an effect (radioactivity for example).

Currie considers that only the case of a zero measurand (absence of effect, θ=0) should be used to establish the decision threshold (Currie, 1999a). This means that he considers the situation where $x$ comes from the background noise (the reference) AND the same for $y$.

$z = x - y$ is thus a realization of the random variable $N = B' - B$ which has the probability distribution

$$p_N(z|0) = \frac{e^{-\frac{(z-0)^2}{4\sigma_B^2}}}{\sqrt{4\pi}\sigma_B} = \frac{e^{-\frac{z^2}{4\sigma_B^2}}}{\sqrt{4\pi}\sigma_B}$$

By setting a threshold of $100\alpha_c\%$, the decision threshold $z_c$, amounts to considering that if we were in the presence only of the reference, only $100\alpha_c\%$ measurements would be higher than this threshold (and would therefore be false positives if we considered them as coming from a signal).

$$\alpha_c = p(N > z_c \,|\, H_0) = \int_{z_c}^{\infty} p_N(z|0)dz$$

(3)

$$\alpha_c = \int_{z_c}^{\infty} p_N(z|0)dz = \frac{1}{\sqrt{2\pi}\sigma_B}\int_{z_c}^{\infty} e^{-\frac{(z-0)^2}{2\sigma_B^2}} dz = 1 - \Phi\left(\frac{z_c}{\sqrt{2}\sigma_B}\right)$$

And thus :

$$z_c = \sqrt{2\,\sigma_B^2}\,\Phi^{-1}(1 - \alpha_c)$$

Where $\Phi$ is the cumulative distribution of the standard normal distribution ($\Phi^{-1}$ is its inverse function or quantille function).

Let us emphasize that for the moment this is one test among other possible ones. Authors have proposed a whole set of tests (Altshuler & Pasternack, 1963; Lehmann & Romano, 2005a; Strom & MacLellan, 2001). However, it is possible to prove that this test is in fact the best possible in the sense of the Neyman-Pearson approach. [Cf Annexe 4].

## 3.3. Detection limit

The detection limit $\theta_d$ in its frequentist definition is the smallest true value of the measurand (parameter) $\theta$ which guarantees the specified probability of being detectable by the measurement method (ISO, 2010a). This will coincide with the greatest true value which would have a non-negligible probability of being considered non-significant by the measurement. It could give rise to observations equal to the decision threshold. In fact, the top of the range of $\theta$ which would have a probability of at least $1 - \beta_c$ to be detected by the measurement method coincides with the bottom of the range of $\theta_d$ having a probability of at most $\beta_c$ to be considered as non-significant. The lowest reasonably detectable value coincides with the highest value likely to be classified as nonsignificant. In Currie's frequentist formulation (Currie, 1968), keeping the previous notations $N$ and $z$, we get :

$$\beta_c = p(N < z_c \,|\, \theta_d) = \int_{-\infty}^{z_c} p_N(z|\theta_d)dz$$

(4)

And

$$1 - \beta_c = p(N < z_c \,|\, \theta_d) = \int_{z_c}^{\infty} p_N(z|\theta_d)dz$$

This corresponds to finding in the alternative hypothesis $H_a$ the smallest valuer $\theta_d$ of $\theta$ for which we will have a probability $\beta_c$ to have measurements below the decision threshold (false negative). We can rewrite this formula in the form:

$$\beta_c = \int_{-\infty}^{z_c} p_N(x|y, \theta_d)dz$$

We are looking for the value of the parameter $\theta_d$ such that the dispersion of the measurements only very improbable gives measurements below the decision threshold. This supposes that we place ourselves in the case of an alternative hypothesis. ($\theta_d \neq 0$) and therefore, within the framework of the Neyman-Pearson approach. If there is no alternative hypothesis, there is no detection limit (Lehmann, 1993).

Setting $\sigma_N = \sqrt{\sigma_S^2 + 2\sigma_B^2}$, the detection limit is therefore calculated as follows for homoscedastic Gaussians

$$\beta_c = \int_{-\infty}^{z_c} p_N(z|\theta_d)dz = \frac{1}{\sqrt{2\pi}\sigma_N} \int_{-\infty}^{z_c} e^{-\frac{(z-\theta_d)^2}{2\sigma_N^2}} dz = \Phi(\frac{z_c - \theta_d}{\sigma_N}) = 1 - \Phi(\frac{\theta_d - z_c}{\sigma_N})$$

$$\theta_d - z_c = \sigma_N \Phi^{-1}(1 - \beta_c)$$

$$\theta_d = z_c + \sigma_N \Phi^{-1}(1 - \beta_c)$$

It is common practice to assume that $\beta_c = \alpha_c$.

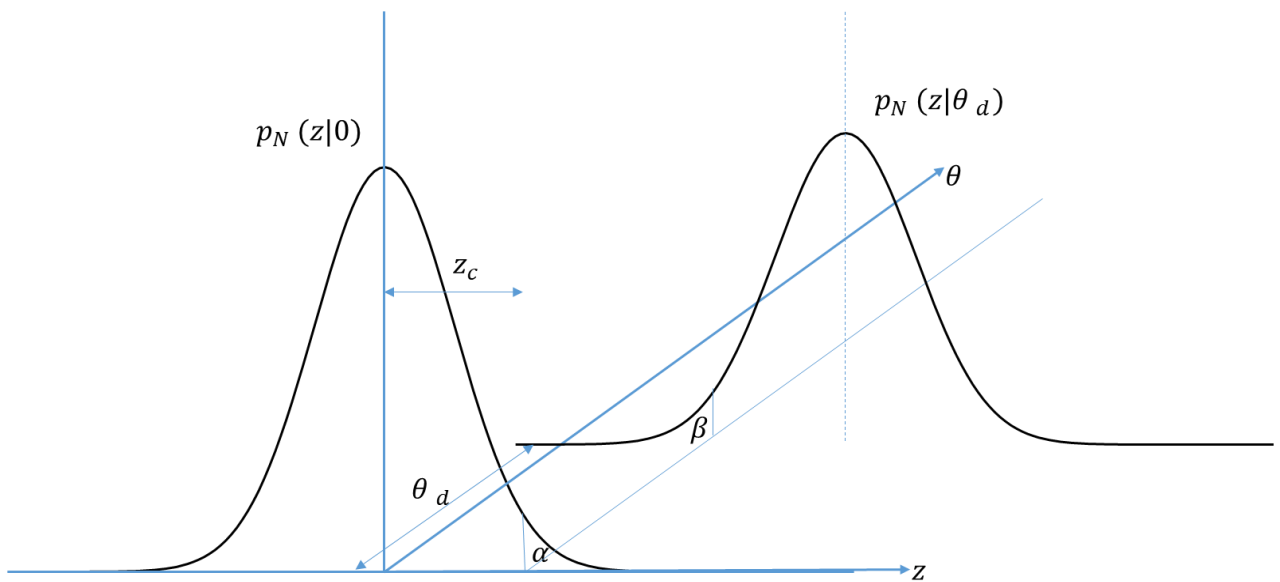

**Figure 2 - Schematic diagram (in 3D cavalier perspective) of the frequentist determination of the detection limit**

## 3.4. Confidence intervals and hypothesis testing

If we use homoscedastic Gaussian distributions for the baseline and sample, we have seen that the distribution of the difference (with the above notations for the random variable $N$) will also be a Gaussian. Using the previous notations, we wish in this section to obtain a confidence interval of the mean $\theta$ of the random variable $N$ modeling the difference between the measurement $G = S + B'$ and the reference $B$.

Consider a series of n measurements of differences between baseline and sample (realizations of the random variable $N$). The average will be $\bar{N}$ and its standard error $\frac{\sigma_N}{\sqrt{n}}$. If we define $t = \sqrt{n}\frac{\bar{N}-\theta}{\sigma_N}$, having a standard normal distribution. $p(t) = \frac{1}{\sqrt{2\pi}} e^{-\frac{t^2}{2}}$.

Using the properties of this distribution, we can set a probability value $\gamma$ such that there exists a $k$ verifiying :
$$P(-k \leq t \leq k) = \gamma$$

$$P\left(-k \leq \sqrt{n}\frac{\bar{N}-\theta}{\sigma_N} \leq k\right) = \gamma$$

$$P\left(-k\frac{\sigma_N}{\sqrt{n}} \leq \bar{N}-\theta \leq k\frac{\sigma_N}{\sqrt{n}}\right) = \gamma$$

$$P\left(\bar{N} - k\frac{\sigma_N}{\sqrt{n}} \leq \theta \leq \bar{N} + k\frac{\sigma_N}{\sqrt{n}}\right) = \gamma$$

We then call t a pivotal quantity

This therefore means that the probability that the interval $\bar{N} \mp k\frac{\sigma_N}{\sqrt{n}}$ includes the parameter $\theta$ (the true value sought) is $\gamma$.

It is important to remember that the value $z$ is a realization of a random variable in the frequentist paradigm used here. $\theta$ is set and must not be considered as a random quantity. It is therefore not possible to use the term probability to talk about $\theta$.

The fact that $z$ (result of a measurement process) is a realization of a random variable is therefore expressed in the following form:

If I repeat my measurement 100 times, in $100\gamma\%$ of the time, the true value of my parameter should be within the different interval calculated for each measurement.

This therefore absolutely does not say that for a particular interval calculated from a measurement $z$, we have $100\gamma\%$ to have the parameter included in this particular interval. It rather specifies that $100\gamma\%$ of my calculated intervals will include the true value $\theta$.

The figure below schematizes the idealized process of this virtual measurement repetition (virtual because these measurement replications are never done in reality). Each point represents a measurement of a sample whose true value is 0.15. The confidence interval would then be determined for each measurement, say with a confidence index of 5%. Only 5 confidence intervals out of 100 will not include the true value. We speak of "coverage probability". This probability is an essential frequentist parameter in statistics and constitutes an important evaluation criterion of the methods. This is also a justification by particle physicists for the use of confidence intervals including a negative part. An interval of the type [-a,b] will have an probability of $\gamma$%. It can therefore includes the value zero and cannot be considered as "unphysical".

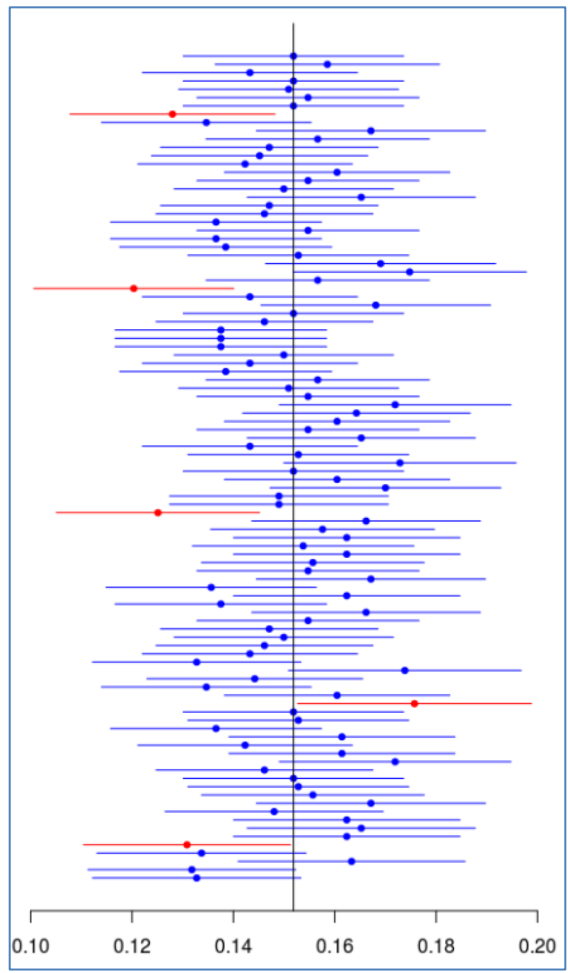

**Figure 3 - Representation of a 100-fold replication of measurement and determination of confidence intervals**

## 3.5. Relation between bounds of the confidence interval and decision threshold

In the case where we want to know wether a measurement is significant, it is enough to check that the zero value is not in the confidence interval calculated from this measurement (Willink, 2006).

**The decision threshold is therefore the smallest measurement for which the confidence interval contains the value zero.**

Let us consider the case of homoscedastic Gaussians.

$$P\left(z - k\frac{\sigma_N}{\sqrt{n}} \leq \theta \leq z + k\frac{\sigma_N}{\sqrt{n}}\right) = \gamma$$

With $k = \Phi^{-1}(\frac{1-\gamma}{2})$ and $\gamma$ the coverage probability of the confidence interval.

For a realization z of the random variable $N$, the lower bound of the confidence interval of $\theta$ is

$z - \Phi^{-1}(\frac{1-\gamma}{2}) \frac{\sigma_N}{\sqrt{n}}$

Thus:

$$z_c - \Phi^{-1}\left(\frac{1-\gamma}{2}\right)\frac{\sigma_N}{\sqrt{n}} = 0$$

$$z_c = \Phi^{-1}\left(\frac{1-\gamma}{2}\right)\frac{\sigma_N}{\sqrt{n}}$$

$z_c = \sqrt{2\sigma_B^2}\, \Phi^{-1}(\frac{1-\gamma}{2})$ in the hypothesise $H_0 = \{\theta = 0\}$

where $\Phi$ is the cumulative distribution function of the gaussian distribution. taking $\frac{1-\gamma}{2} = \alpha_c$, we obtain exactly the same result as with the direct hypothesis test.

$$z_c = \sqrt{2\sigma_B^2}\, \Phi^{-1}(\alpha)$$

We therefore find the Currie decision threshold.

**For the detection limit, the largest parameter value must be determined in the confidence interval compatible with a measurement $z = z_c$.**

**We immediately recognize that the upper bound of the confidence interval of $\theta$ for an observation $z_c$ will give us back the expression of the detection limit.**

More precisely:

$$\theta_d = z_c + \sqrt{2\sigma_B^2}\,\Phi^{-1}(\alpha_c) = 2z_c$$

Here again, we obviously find the result of Currie's approach.

Once we have made a measurement and determined a confidence interval, it is therefore not necessary to also carry out a hypothesis test. Just look if the confidence interval contains 0

In the case of a non-significant result ($z < z_c$), the upper bound of the confidence interval will be lower than the detection limit.

We can schematize everything we have just said as follows:

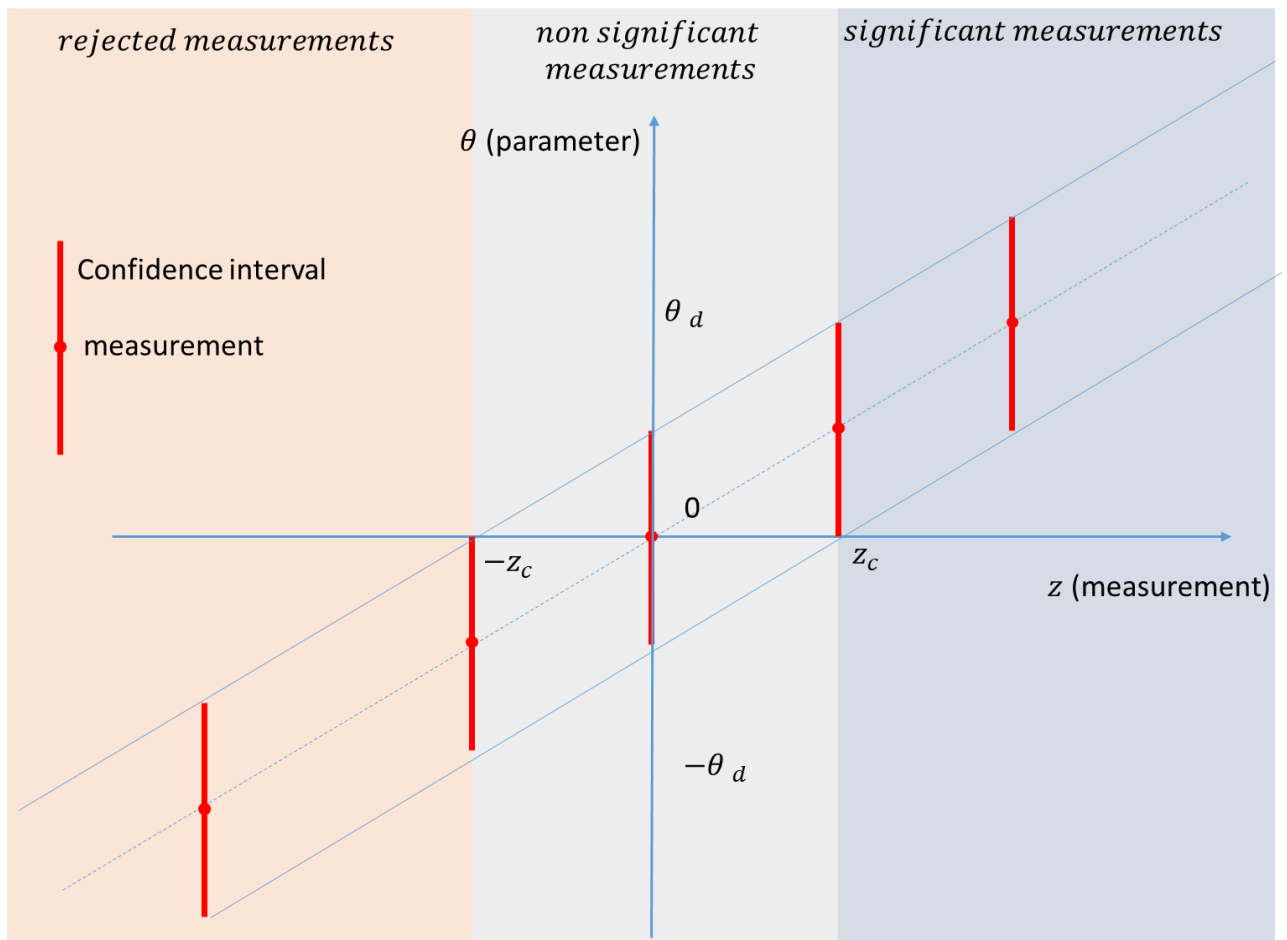

Figure 4 - Representation of the different possible situations during a measurement process

So if we know how to generate the confidence interval, we have all the information to determine the decision threshold and detection limit. If the interval "touches" the zero value of the parameter, the measurement will be equal to the decision threshold. The uncertainty will then be 100% and the upper limit of the interval will then be the detection limit. In addition, the hypothesis test used to determine it is that from Neyman Pearson's lemma, it is therefore the most efficient test.

> **Uncertainty=100% $\Leftrightarrow z = z_c \Leftrightarrow$ non significant results**
>
> **$\Leftrightarrow$ lower bound of the confidence interval =0 $\Leftrightarrow$ upper bound=detection limit $\theta_d$**

If the interval includes a negative part and a positive part then its uncertainty will be greater than 100%. The hypothesis test on the value 0 cannot be rejected.

> **Uncertainty > 100% $\Leftrightarrow$ measurement < decision threshold $\Leftrightarrow$ non significance**
>
> **$\Leftrightarrow$ lower bound of the confidence interval <0 $\Leftrightarrow$ upperbound<detection limit $\theta_d$**

It is even possible to add another consideration. If the sample measurement is too low compared to the baseline, then it can be considered that it is no longer reasonably compatible with the baseline estimate. A measurement that is too negative can no longer be explained by measurement fluctuations. It will then be appropriate to carefully examine this measurement and this sample in order to rule out any possibility of error or bias.

> **measurement < rejection threshold=$-z_c \Leftrightarrow$ suspicious results**
>
> **$\Leftrightarrow$ upperbound<0**

We could therefore talk about a rejection threshold.

Let us emphasize once again that we are reasoning here on the variable $N$ (net variable) and not on $S$ (the variable sought). The first can absolutely have a confidence interval with a negative part. This confidence interval is determined using the probability distribution of $N$. We cannot determine the confidence interval for $S$ because we do not have its probability distribution in the absence of the exact noise contribution within sample measurement.

# 4. BAYESIAN APPROACH FOR HOMOSCEDASTIC GAUSSIANS

## 4.1. Introduction to Bayesian methods

Bayesian statistics provide a way to infer desired physical parameters from observational data. The "classic" method (although subsequent to the birth of Bayesian methods), called frequentist, assumes that we are looking for an unknown but fixed parameter.The bayesian method (Gelman et al., 2013) assumes that the relationship between the observed quantities and the parameters is statistical. Mathematically, this amounts to considering the parameters of interest as random variables with a probability density intended to completely describe our beliefs or knowledge about it.

The parameters are therefore modeled in terms of probability distributions: Starting from "a priori" distributions on these parameters, they are updated according to observations to produce so-called "a posteriori" distributions. Note that there are numerous mathematical-epistemological interpretations of Bayesian methods. Some authors have even counted 46,656 possible varieties of "Bayesianism"(Good, 1976).

Because we assign a distribution to the parameters, statistical inference is reduced to the application of probability theory.

If we consider a joint probabilistic distribution $p_{X,U}(x, \mu)$ of observations and parameters, $X$ and $U$ respectively ,it is possible to write $p_{XU}(x, \mu) = p_X(x|\mu)p_U(\mu)$.

The first term $p_X(x|\mu)$ (probability of having an observation $x$ knowing the parameter $\mu$) is called the likelihood. It will correspond to the chosenstatistical model (Gaussian for example). This concept of likelihood also exists in frequentist methodologies and corresponds to the modeling of the distribution of observations as a function of parameters (Gaussian distribution of mean $\mu$ for example).

The second term $p_U(\mu)$ is the a priori or prior distribution of the parameter. It quantifies our prior beliefs about the distribution of parameters even before taking into account observations.

Likewise, we can write $p_{XU}(x, \mu) = p_U(\mu|x)p_X(x)$. Bayes' theorem simply consists of writing :

$$p_U(\mu|x) = \frac{p_X(x|\mu)p_U(\mu)}{p_X(x)}$$

Where $p_X(x)$ is the probability of the observation $x$ integrated on all possible values of $\mu$, sometimes called marginal likelihood. $p_U(\mu|x)$ is the a posteriori distribution (after the observations), the posterior. We will say that the posterior is equal to the product of the likelihood and the prior divided by the marginal likelihood.

Given that the posterior is a probability density which must be normalized to 1, we can consider $p_x(x)$ c as a simple normalization constant (because it does not depend on µ i.e $p_X(x) = \int_{-\infty}^{\infty} p_{XU}(x, \mu)d\mu = \int_{-\infty}^{\infty} p_X(x|\mu)p_U(\mu)d\mu$ ):

$$p_U(\mu|x) \sim p_U(x|\mu)p_U(\mu)$$

The posterior is proportional to the product of the prior and the likelihood. We carried out a Bayesian inversion.

Subsequently, a prior of a parameter µ will be denoted in the form π(µ) to facilitate understanding. A crucial point of Bayesian methodology, which is also the cornerstone of the criticisms addressed to it, is its dependence on priors. There are many ways to choose them which are the subject of fierce and heated discussion.

### 4.1.1. Chosing a prior

A crucial point of Bayesian methodology is its dependence on priors. The choices of priors can be motivated by past experiences or by intuition, but also by computational aspects as in the case of conjugate priors.

Aware of this dependence on priors, in addition to the absence of a priori knowledge in a number of problems, numerous works have been interested in the definition of "non-informative" priors whose influence on the posterior probability is reduced to a minimum.

This type of prior respects the so-called Jeffreys rule in connection with invariances by transformation (translation in this case). With this prior, the confidence and credibility intervals often coincide (Jaynes, 1968; Karlen, 2002; Rosenkrantz, 1989; Severini, 1991) and it is therefore possible to use frequentist and Bayesian concepts interchangeably. In particular, the coverage probabilities will naturally apply to the credibility intervals and it will be possible to give a probability for a value of the parameter.

In the rest of the document, we will essentially use non-informative priors to investigate the questions asked in the Bayesian framework.

## 4.2. Presentation of the problem

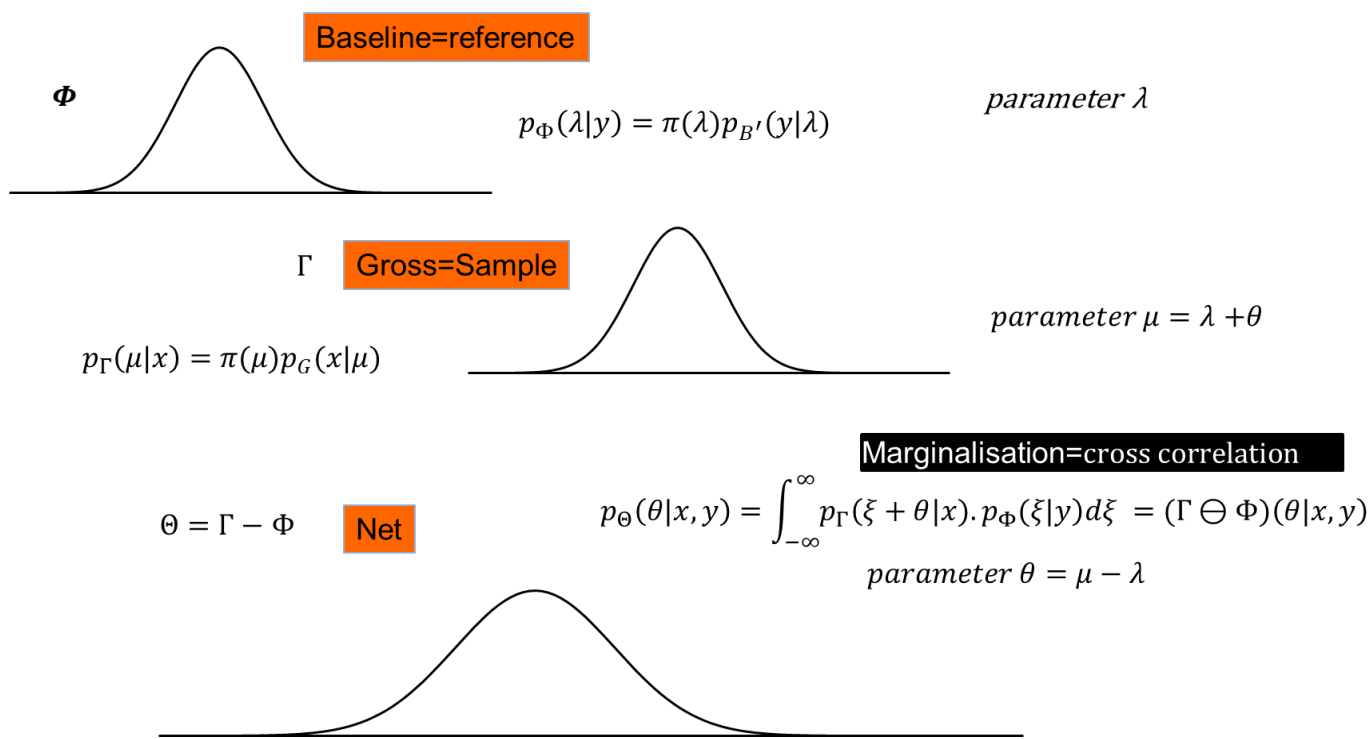

Figure 5 - Bayesian schematic diagram of the problem

As with the previous frequentist chapter, the variable of interest in this section is the random variable $N$ (modeling the net measurement).

We wish to obtain its posterior distribution and exploit the credibility intervals.

As we have already mentioned, the frequentist point of view is to consider that observations are realizations of random variables whose probability distributions have fixed but unknown parameters.

In the Bayesian paradigm, models supposed to account for observations can incorporate constraints on the parameters (more precisely on the information available on the parameters), treated as random variables. In particular, for random variables $\Gamma$ et $\Phi$ modeling the sample measurement and the reference respectively, we can determine the following probability densities:

$$p_\Gamma(\mu|x) = \pi(\mu)p_G(x|\mu)$$
$$p_\Phi(\lambda|y) = \pi(\lambda)p_B(y|\lambda)$$

where $\pi(\mu)$ et $\pi(\lambda)$ are the corresponding priors.

Our choice in the present study is to consider non-informative priors $\pi(\mu) = \pi(\lambda) = 1$ :

$$p_\Gamma(\mu|x) = \pi(\mu)p_G(x|\mu) = \frac{e^{-\frac{(x-\mu)^2}{2\sigma^2}}}{\sqrt{2\pi}\sigma}$$

$$p_\Phi(\lambda|y) = \pi(\lambda)p_B(y|\lambda) = \frac{e^{-\frac{(y-\lambda)^2}{2\sigma^2}}}{\sqrt{2\pi}\sigma}$$

The joint probability density is therefore :

$$p_{\Gamma\Phi}(\lambda,\mu|x,y) = \pi(\mu)\pi(\lambda)p_\Gamma(\mu|x)p_\Phi(\lambda|y) = \frac{e^{-\frac{(x-\mu)^2}{2\sigma^2}}}{\sqrt{2\pi}\sigma}\frac{e^{-\frac{(y-\lambda)^2}{2\sigma^2}}}{\sqrt{2\pi}\sigma}$$

We know that if $G$, $B'$ are $S$ are gaussian densities of means $\mu$, $\lambda$ and $\theta$ respectively, then by construction of $G = S + B'$ we have $\mu = \lambda + \theta$ and by reparametrizing the equation above we find:

$$p_{\Gamma\Phi}(\lambda,\theta|x,y) = \pi(\lambda+\theta)\pi(\lambda)p_\Gamma(\lambda+\theta|x)p_\Phi(\lambda|y) = \frac{e^{-\frac{(x-\theta)^2}{2\sigma^2}}}{\sqrt{2\pi}\sigma}\frac{e^{-\frac{(y-\lambda)^2}{2\sigma^2}}}{\sqrt{2\pi}\sigma}$$

The Bayesian approach to getting rid of a nuisance parameter like $\lambda$ est simply to integrate with respect to this parameter. This is called a marginalization.

$$p_\Theta(\theta|x,y) = \int p_{\Gamma\Phi}(\lambda,\theta|x,y)d\lambda = \int \frac{e^{-\frac{(x-\lambda-\theta)^2}{2\sigma^2}}}{\sqrt{2\pi}\sigma}\frac{e^{-\frac{(y-\lambda)^2}{2\sigma^2}}}{\sqrt{2\pi}\sigma}d\lambda = \frac{e^{-\frac{(z-\theta)^2}{4\sigma^2}}}{\sqrt{4\pi}\sigma}$$

With $z = x - y$. Given that we integrate over the entire domain of definition of $\lambda$, $\theta$ will take all possible values of $]-\infty, +\infty[$.

## 4.3. Credibility interval and hypothesis testing

In the Bayesian approach, the posterior distribution probabilistically contains all the information on the parameter. Given an observation z, the credibility interval $[a,b]$ à $(1-\gamma)100\%$ is defined by :

$$p(a \leq \theta \leq b|z) = (1-\gamma)$$

Unlike confidence intervals, it is therefore legitimate here to speak of a probability for the parameter to be found in this interval (Bolstad, 2007).

**There are several ways to perform hypothesis testing or their equivalent in the Bayesian approach.**

**We will use Bolstad's suggestion here (Bolstad, 2007). We will test the credibility of this null hypothesis by examining whether the null value is included in the credibility interval. If this is not the case, we will reject this hypothesis. Otherwise, we will consider this to be a credible value.**

The approach is essentially the same as examining whether the zero value is part of the confidence interval by replacing the latter with the credibility interval. The main advantage is to eliminate the one-off nature of the test. We do not test if the parameter is equal to a precise value (which from a mathematical point of view is a set of zero Lebesgues measure) but wether this parameter is included in an interval.

This agrees with the point of view of a certain number of statisticians who consider that the gap between hypothesis testing and estimation is not necessary. (Bolstad, 2007; Cumming, 2014; Cumming & Calin-Jageman, 2016; Kruschke & Liddell, 2018). It is enough to know the posterior distribution of the parameter of interest to both estimate the parameter and carry out hypothesis tests. Some of these authors even speak of "new statistics".As we will see, knowledge of the posterior distribution is sufficient for both estimation and hypothesis testing corresponding to the decision threshold in the field of low-level metrology..

### 4.3.1. Decision threshold

In a similar way to 3.2, we propose to define a Bayesian decision threshold.

The decision threshold would be the smallest net measurement value $z_c$ such that the zero value is the lower limit of the credibility interval:

$$\alpha_c = \int_{-\infty}^{0} p_\Omega(\theta|z_c)d\theta = P(\Omega < 0|z_c)$$

In the case of members of the "location family" to which the Gaussians belong, we know that $p_N(z|\theta) = f(\theta - z) = p_\Omega(\theta|z)$ because we then take a non-informative prior. This implies that, by change of variable $t = \theta - z$, we can write:

$$\alpha_c = \int_{z_c}^{\infty} p_N(z|0)dz = \int_{-\infty}^{0} p_\Theta(t|z_c)dt$$

The frequentist and Bayesian decision thresholds therefore coincide.Note that this definition of the decision threshold based on the inclusion of zero in the credibility interval corresponds exactly to that suggested by Jaynes (Jaynes, 1968, 2003). Indeed, Jaynes proposed using decision thresholds based on probability in this type of case. $\alpha$ to have a given sign (Jaynes, 1968, 2003). Thus, we will determine a threshold $z_c$ such that:

$$\alpha_c = p_{G,B}(\mu < \lambda|z_c)$$

By definition, given that $\theta = \mu - \lambda$ where $\lambda$ is the position parameter of the sample et µ that of the reference:

$$p_{G,B}(\mu < \lambda|z_c) = p_\Theta(\theta < 0|z_c) = \int_{-\infty}^{0} p_\Theta(\theta|z_c)d\theta$$

*(5)*

We therefore seek to determine an observed value $z_c$ which will ensure that the probability for $\theta$ to be of negative sign is equal to a given $\alpha_c$% (for example 2,5%). %). Remember that, by definition, this corresponds to a probability $\alpha_c$ that the value of the sample parameter is lower than that of the reference. This corresponds to having a hypothesis $H_0 : \theta < 0$ and an alternate hypothesis $H_1 : \theta > 0$. This is the exact Bayesian correspondence of the decision threshold defined in 3.2. To our knowledge, only Lemay has explicitly used this criterion in the field of metrology (Lemay, 2012, 2015). Others have done it in a somewhat concealed or unconscious way by firstly excluding the possibility of negative values of $\theta$ then somewhat paradoxically considering despite everything that this represented the null hypothesis (Kirkpatrick & Young, 2009)

Another remark that can be made is that this definition based on a non-informative prior would not make sense if we had prohibited the existence of negative values (with a prior of the type for example $\pi(\theta) = 1 \; if \; (\theta > 0)$).

### 4.3.2. Detection limit

Similarly, we can determine a Bayesian detection limit. We are looking for the greatest possible value of the parameter compatible with the measurement of the decision threshold.

We look for the upper bound of the credibility interval such that:

$$\beta_c = \mathrm{P}(\Theta > \theta_d | z_c) = \int_{\theta_d}^{\infty} p_\Theta(\theta|z_c)d\theta = \int_{-\infty}^{z_c} p_N(z|\theta_d)dz$$

The Bayesian detection limit will therefore coincide with the frequentist detection limit. It is therefore possible for us in this case to move indifferently from one point of view to another and to use the different insights provided.

## 5. SYNTHESIS

## 5.1. Frequentist point of view

### 5.1.1. Calculation of the decision threshold and detection limit

As defined previously, homoscedasticity is the property of keeping a constant variance. In the case that concerns us, this means that the sample and the reference have the same variance. In other words, the signal has zero variance (adding a signal to the noise is done without increasing the uncertainty).

$$\sigma_S^2 = 0$$

$$\sigma_N = \sqrt{\sigma_B^2 + \sigma_G^2} = \sqrt{2\,\sigma_B^2 + \sigma_S^2} = \sqrt{2\,\sigma_B^2}$$

Apart from pathological cases, homoscedasticity can be modeled by the Dirac distribution.

$$p_S(z|\theta) = \delta(z - \theta)$$

In the absence of measurement uncertainty, there is no intrinsic variability of the signal

$$p_N(z|\theta) = \int_{-\infty}^{\infty} p_S(z-w|\theta)p_N(w|0,2\sigma_B^2)dw = \int_{-\infty}^{\infty} \delta(z-w-\theta)\frac{e^{-\frac{w^2}{4\sigma_B^2}}}{\sqrt{4\pi}\sigma_B}dw = \frac{e^{-\frac{(z-\theta)^2}{2\sigma_N^2}}}{\sqrt{2\pi}\sigma_N}$$

We can thus calculate the decision threshold $z_c$ in accordance with the method set out in 3.2. It must be such that, fo r agiven $\alpha_c$ :

$$\alpha_c = \int_{z_c}^{\infty} p_N^{H_0}(z|0)dz = \frac{1}{\sqrt{2\pi}\sigma_N}\int_{z_c}^{\infty} e^{-\frac{(z-0)^2}{2\sigma_N^2}} dz = 1 - \Phi(\frac{z_c}{\sigma_N})$$

And so :

$$z_c = \sigma_N \Phi^{-1}(1-\alpha_c) = \sqrt{2\sigma_B^2}\Phi^{-1}(1-\alpha_c) = \sigma_N \Phi^{-1}(1-\alpha_c)$$

(6)

The detection limit is calculated as explained in 3.3 :

$$\beta_c = \int_{-\infty}^{z_c} p_N(z|\theta_d)dz = \frac{1}{\sqrt{2\pi}\sigma_N}\int_{-\infty}^{z_c} e^{-\frac{(z-\theta_d)^2}{2\sigma_N^2}} dz = \Phi(\frac{z_c - \theta_d}{\sigma_N}) = 1 - \Phi(\frac{\theta_d - z_c}{\sigma_N})$$

$$\theta_d - z_c = \sigma_N \Phi^{-1}(1-\beta_c)$$

$$\theta_d = z_c + \sigma_N \Phi^{-1}(1-\beta_c)$$

If we set $\alpha_c = \beta_c$, then :

$$\theta_d = 2z_c$$

(7)

This is the result of the classic Currie approach (Currie, 1968) for homoscedastic distributions.

In general, the results are returned as follows:

| measurements | estimation |
| --- | --- |
| $z < z_c$ (non significants) | $< \theta_d$ |
| $z > z_c$ (significants) | $\theta \pm \delta$ |

where $\delta$ is the uncertainty.

### 5.1.2. Comparison with confidence interval

As we indicated in the paragraph 3.5, just knowing the confidence interval [a,b] is enough to know whether a result is significant or not:

| | |
| --- | --- |
| $0 \in [a, b]$ | **Non significant results (0 is aplausible value)** |
| $0 \notin [a, b]$ | Significant results (0 is not a plausible value) |

It is therefore not even necessary to calculate the decision threshold. The detection limit can be useful, in the sense that, in the absence of measurement of the sample, it makes it possible to estimate the value of the parameter which would be the smallest possible while still being reasonably likely to be detected. It gives an idea of the performance of the measurement method.

On the other hand, determining the confidence interval gives for each specific measurement an upper limit to the set of parameter values compatible with this measurement. Instead of having a limit valid for all non-significant results, with the confidence interval we have a limit specific to each measurement (therefore more precise).

Thus, if from a measurement $z$ we determine a confidence interval for the confidence index $\alpha$:

$$[\theta^-(z), \theta^+(z)]$$

If $\theta^-(z) \leq 0$ the result is not significant and we can then estimate that the parameter (of the interval) $\theta < \theta^+(z)$ avec $\theta^+(z) \leq \theta_d$

## 5.2. Bayesian point of view

The Gaussian distribution is a member of the family location familly. As mentioned previously, we will therefore choose a uniform prior for reasons of symmetry (so-called "non-informative" prior) as is customary (Box & Tiao, 1973). At this point, the credibility intervals exactly coincide with the confidence intervals (Jaynes, 2003; Jaynes & Kempthorne, 1976). We will therefore obtain exactly the same results as with the frequentist method.

If $[a, b]$ is the credibility interval, we will deduce the same type of consideration as with the frequentist confidence interval.

| $0 \in [a, b]$ | Non significant results (0 credible value) |
| $0 \notin [a, b]$ | Significant results (0 is not a credible value) |

The confidence intervals here are strictly equal to the credibility intervals. So whether it is classic hypothesis testing, the confidence interval criterion or the credibility interval, the results will be the same for statistical significance, decision thresholds or detection limits. In particular, we will have the relationship:

$$\theta_d = 2z_c$$

(8)

This implies that the probability of recovery will also necessarily be adequate for the credibility interval. Indeed, the confidence interval is constructed from the principle of the probability of recovery. But the credibility interval coincides with the confidence interval for the case of homoscedastic Gaussians (Karlen, 2002). Therefore, the coverage probability of the Bayesian credibility interval will be the same as that of the confidence interval. Statistical performance should therefore be adequate.

## 5.3. Verification by simulation

Voigtman underlines the crucial importance of verifying the statistical properties of the quantities that we calculate (Voigtman, 2017) :

*« Finally, computer simulations are absolutely essential; no one who has seriously studied the fundamental aspects of detection limits has had infallible intuition, most certainly including the author. Indeed, in regard to detection limit theory and practice, it is fair to say that competently devised and performed computer simulations are the most effective way, by far, to avoid fooling oneself. »( page 197°*

He carried out simulations to check that the characteristic limits determined by Currie had the correct statistical properties. In this case, he verified for homoskedastic ideal systems that by setting the value of $\alpha_c$ and deducing the decision thresholds and detection limits, we obtain $100.\alpha_c\%$ false positives by simulations ( 5,005 ± 0.033% compared to a theoretical rate of 5% for 1 million draws) (Voigtman, 2017). Let us specify that here we are indeed in the presence of false positives since the simulation will generate measurements for a zero-parameter distribution. False negative rates were also checked (4,998 ± 0.028% compared to a theoretical rate of 5% with 1 million draws). We can therefore consider that these limits have the desired statistical properties for homoscedastic systems. The confidence interval will by definition have the correct coverage probability since it is built to do so. It would be possible but unnecessary to check it.

## 5.4. Conclusions on the homoscédastic case

So whether for classic hypothesis tests, the confidence interval criterion or the credibility interval, the results will be the same for the statistical significance, the decision thresholds or the detection limits. The credible intervals being identical to to the confidence intervals, it implies that the coverage probability will necessarily be adequate. It is therefore not necessary to carry out a specific hypothesis test in addition since the simple determination of the confidence interval is enough to give us all the necessary information. We are not adding extra work to ourselves by proceeding in this way, we save ourselves work

Furthermore, as mentioned before, the Neyman-Pearson lemma guarantees that this is the best possible test.

From the confidence interval, it is then possible to provide for non-significant results an upper limit to the desired parameter, more precise than the simple detection limit.

# 6. POISSON DISTRIBUTIONS

## 6.1. frequentist approach

The Poisson case corresponds to counting measurements of the reference and the sample, modeled respectively by random variables $B$ et $G = S + B'$ , Poisson distributions of parameters $\lambda$ and $\mu = \theta + \lambda$ :

$$p_B(y|\lambda) = \frac{\lambda^y e^{-\lambda}}{y!} \; et \; p_G(x|\mu) = \frac{\mu^x e^{-\mu}}{x!}$$

The joint probability will then be

$$p_B(x,y|\lambda) = \frac{\lambda^y e^{-\lambda}}{y!} \frac{\mu^x e^{-\mu}}{x!} = \frac{\lambda^y e^{-\lambda}}{y!} \frac{(\theta + \lambda)^x e^{-(\theta+\lambda)}}{x!}$$

As in the previous chapters, the objective is to carry out an inference on the parameter $\theta$ and in particular test the hypothesis $H_1=\{\theta > 0\}$ against a null hypothesis $H_0 = \{\theta = 0\}$ or $H_0 = \{\theta \leq 0\}$

A first natural approach in the spirit of the Gaussian case presented in the previous chapters would be to consider the random variable of the difference $N = G - B$.

Let us mention, however, that the probability distribution of the difference between two Poisson distributions is not a Poisson distribution but a Skellam distribution. (Skellam, 1946) :

$$p_k(x - y | \lambda, \theta) = e^{-(\theta + 2\lambda)} \left( \left| \frac{\lambda}{\lambda + \theta} \right| \right)^{\frac{(x-y)}{2}} I_{|x-y|}(2\sqrt{\lambda(\lambda + \theta)})$$

where $I_{|k|}$ is the modified Bessel Function of the First Kind.

We see that this probability density will depend both on $\theta$ and $\lambda$. We will also note that intrinsically Skellam's law completely authorizes that $\mu < \lambda$ and so $\theta < 0$

Thus and contrary to the Gaussian laws previously studied, the probability density of the difference $N = G - B$ depends on the parameter $\lambda$ which we can consider here as a "nuisance" parameter as opposed to the parameter of interest $\theta$ (Liseo, 2005; Pawitan, 2001).

We cannot therefore directly use the difference $N$ to study $\theta$ without knowing or previously estimating this nuisance parameter $\lambda$. What is done in the standards and reference texts is either to consider $\lambda$ as known (in particle physics or astrophysics for example (Lista, 2016) or to resort to a Gaussian approximation of the Poisson law in order to try to reduce it to the Gaussian case (see previous chapters).

Is it possible in this case to find another joint probability density of $G$ and $B$ not depending on $\lambda$ p for the case of Poisson laws?

Let us formulate the problem in the case of the decision threshold: in its traditional formulation, the decision threshold is the observation $z_c$ of the random variable $N$ such that :

$$P_{H0}(N > z_c) = \alpha$$

As presented above, it is not possible in the Poissonian case (and more generally in the non-Gaussian case, we will come back to this) to determine directly $z_c$ from this equation due to the presence of the nuisance parameter $\lambda$.

We see clearly that the reason for this difficulty is to immediately consider the random variable difference $N$.

We propose to get around this difficulty by considering conditional random variables, recognized as one of the methods for eliminating nuisance parameters. (Basu, 2011; Liseo, 2005; Sprott, 2008).

## 6.2. Conditional likelihood and hypothesis testing

The crucial point of our approach is to consider conditioning by sufficient statistics.

This is a very natural approach in statistics when faced with nuisance parameters. (Sprott, 2008) since, by definition, the conditional probability density of a random variable by its sufficient statistics is independent of the parameter of this law.

In the particular case of Poisson's laws which interest us here, the sufficient statistic is simply the sum of the random variables $B + B'$ (more generally, the sum of random variables is the sufficient statistics of probability laws belonging to the family of natural exponentials).

Consequently, we can generalize in a very simple way the previous definition of the decision threshold by considering the conditional probability by $B + B'$:

$$p_{B'|B+B'}(B' = y'|B + B' = y + y') = \frac{p_{B',(B + B' = y + y', B' = y')}}{p_{B+B'}(B + B' = y + y')} = \frac{p_B(B = y) p_{B'}(B' = y')}{p_{B+B'}(B + B' = y + y')}$$

$$= \frac{\frac{\lambda^y e^{-\lambda}}{y!} \frac{\lambda^{y'} e^{-\lambda}}{y'!}}{\frac{(2\lambda)^{y+y'} e^{-2\lambda}}{(y + y')!}}$$

$$p_{B'|B+B'}(B' = y'|B + B' = y + y') = \frac{(y + y')!}{y! \, y'!} \left(\frac{1}{2}\right)^{y+y'}$$

This expression is that of a negative binomial law for $y'$ with parameters $y$ and $\frac{1}{2}$.

We therefore have a probability density of $y'$ containing no unknown parameters and, in particular, no $\lambda$. We have eliminated the nuisance parameter.

In our situation, assuming $H_0 = \{\theta = 0\}$, there is no signal ($S = 0$ and $G = B'$). If we make a measurement $y$ for the reference ($B$), then we will have as probability density of $x$ in the sample:

$$p_{B'|B+B'}(G = x|B + B' = y + x) = \frac{(y + x)!}{y! \, x!} \left(\frac{1}{2}\right)^{y+x}$$

Let's lighten the notations by noting $p_c(x; y) = p_{B'|B+B'}(G = x|B + B' = y + x) = \frac{(y+x)!}{y!x!} \left(\frac{1}{2}\right)^{y+x}$.

$p_c(x; y)$ is indeed a probability density of $x$ when we have a measurement of $y$ and without any unknown parameters.

We know that the cumulative distribution function of the negative binomial distribution of parameters $y$ and $\frac{1}{2}$ is (JOHNSON et al., s. d.) the regularized incomplete beta function :

$$p(x < x_c; y) = I_{\frac{1}{2}}(y + 1, x_c + 1) = 1 - I_{\frac{1}{2}}(x_c + 1, y + 1)$$

where $I_x(a, b) = \frac{B_x(a,b)}{B(a,b)}$ is the regularized incomplete beta function with $B_x(a, b) = \int_0^x \omega^{a-1}(1 - \omega)^{b-1} d\omega$ the regularized incomplete beta function.

We then have:

$$p(x > x_c; y) = I_{\frac{1}{2}}(x_c + 1, y + 1)$$

For a given level of confidence $\alpha_c$ ($100.\alpha_c\%$ is the fixed false positive rate that we do not wish to exceed), the decision threshold $z_c = x_c - y$ can be defined by:

$$I_{\frac{1}{2}}(x_c + 1, y + 1) = \alpha_c$$

*(9)*

## 6.2.1. Uniformly most powerful test

We know that for the family of natural exponentials, there always exist uniformly more powerful hypothesis tests (UMP tests,-see corollaire 3.4.1in (Lehmann & Romano, 2005b)). Consider the joint probability density

$$p_{G,B}(G = x, B = y) = p_G(G = x|\mu)p_B(B = y|\lambda) = \frac{\mu^x}{x!}e^{-\mu}\frac{\lambda^y}{y!}e^{-\lambda} = \frac{e^{-\lambda-\mu}}{x!\,y!}e^{y\,ln(\lambda)+x\,ln(\mu)}$$

$$= \frac{e^{-\lambda-\mu}}{x!\,y!}e^{(y+x)\,ln(\lambda)+(x\,ln(\mu)-x\,ln(\lambda))} = \frac{e^{-\lambda-\mu}}{x!\,y!}e^{\left(x\,ln\left(\frac{\mu}{\lambda}\right)+(y+x)\,ln(\lambda)\right)}$$

Setting $\theta = \ln\left(\frac{\mu}{\lambda}\right)$, we get :

$$p_{G,B}(G = x, B = y) = \frac{e^{-\lambda-\mu}}{x!\,y!}e^{(x\theta+(y+x)\,ln(\lambda))}$$

From theorem 4.1.1 of Lehmann (Lehmann & Romano, 2005b) the UMP test to decide wether $\theta > 0$ is based upon the test statistic $p(G = B' = x|B' + B = x + y)$.

The hypothesis test we used to define the decision threshold $z_c = x_c - y$, is the UMP test for this hypothesis. The decision threshold $z_c$ is optimal. This therefore confirms the interest of conditional likelihoods in the presence of nuisance parameters.

## 6.2.2. Conditional likelihood in the presence of a signal

Now suppose the presence in the sample of a signal $S$, obeying a Poisson law of parameter $\theta$. After a measurement $y$ of the reference, we will have a $y'$ contribution of the noise in the sample. The probability density of $y'$ will be

$$p_{B'|B+B'}(B' = y'|B + B' = y + y')$$

If we have a measurement of $x$ for the sample, we have to substract the contribution of the noise to know the contribution of the signal:

$$P(S = x - y', B' = y'|\theta, B + B' = y + y') = \frac{P(S = x - y', B' = y', B + B' = y + y'|\theta)}{P(B + B' = y + y')} =$$

$$= \frac{P(S=x-y',B'=y',B=y|\theta)}{P(B+B'=y+y')} = \frac{P(S=x-y'|\theta),P(B'=y'|B+B'=y'+y)P(B+B'=y+y')}{P(B+B'=y+y')}$$

$$= P(S = x - y'|\theta)\,P(B' = y'|B + B' = y' + y) = P(S = x - y'|\theta)p(y'; y)$$

If we do the sum over all possible $y'$, we get the following density

$$p_c(x,y|\theta) = \sum_{y'=0}^{x} P(S = x - y'|\theta)p(y';y)$$

As a convolution of a negative binomial $p_c(y';y)$ and a Poisson distribution $P(S = x - y'|\theta)$. This simply expresses the idea that the probability of having a measurement $x$ for the sample will be the sum of the probabiility of getting $x - y'$ from the Poisson distribution times the probability of measuring $y'$ knowing that we measured $y$ for the reference.

This convolution of a Poisson distribution and a negative binomial distribution is what is called a Delaporte distribution. (JOHNSON et al., s. d.). This distribution has no simple expression but can be evaluated numerically (a R « package » exists). Note that $p_{G|B}(x|y,\theta)$ will have a mean of $\theta + y$ and a variance of $\theta + 2y$.

## 6.3. Bayesian method

Let us first recall that a gamma law $Gamma(\alpha, \beta)$ pcan be expressed as :

$$\Gamma(x|\alpha, \beta) = \frac{x^{\alpha-1} e^{-\beta x} \beta^\alpha}{\Gamma(\alpha)}$$

To directly eliminate the nuisance parameter in a Bayesian approach, it would be necessary to proceed by marginalization (integrating on $\lambda$ as we did for the homoscédastic gaussians). Before that, we will determine the posterior distribution of the parameter of interest. Using the notations from the previous paragraph, we know that the joint probability is:

$$p_{(G,B)}(x,y|\mu,\lambda) = \frac{\mu^x e^{-\mu}}{x!} \frac{\lambda^y e^{-\lambda}}{y!}$$

Using the Bayesian formalism with a generalized Jeffreys prior (Jeffreys, 1946) $\pi(\mu, \lambda) = \frac{1}{(\lambda\mu)^a}$, where the hyperparameter $a = 0$ for a uniform prior, $a = 1/2$ a Jeffreys prior and $a = 1$ for an inverse prior. This will allow us to evaluate the influence and adequacy of the prior at the end of the calculation. This is good practice in the application of Bayesian methodologies. We can obtain the joint probability density of the parameters:

$$p_{(\Gamma,\Phi)}(\mu,\lambda|x,y) \sim \frac{\mu^x e^{-\mu}}{x!} \frac{\lambda^y e^{-\lambda}}{y!} \pi(\mu,\lambda) = \frac{\mu^{x-a} e^{-\mu}}{x!} \frac{\lambda^{y-a} e^{-\lambda}}{y!}$$

Defining $y_a = y - a$ and $x_a = x - a$, the joint probability density of the parameters (posterior distribution) would then be:

$$p_{(\Gamma,\Phi)}(\mu,\lambda|,y_a) = C\,\pi(\lambda,\mu) \frac{\lambda^y e^{-\lambda}}{y!} \frac{\mu^x e^{-\mu}}{x!} = C \frac{\lambda^{y_a} \mu^{x_a} e^{-(\lambda+\mu)}}{y!\,x!}$$

With C a normalization constant. We can also prove that this constant is $C = \frac{x!\,y!}{\Gamma(x_a+1)\Gamma(y_a+1)}$.

$$p_{(\Gamma,\Phi)}(\mu,\lambda|x_a,y_a) = \frac{\lambda^{y_a}\mu^{x_a}e^{-(\lambda+\mu)}}{\Gamma(y_a+1)\Gamma(x_a+1)} = \frac{\lambda^{y_a}(\lambda+\theta)^{x_a}e^{-(2\lambda+\theta)}}{\Gamma(y_a+1)\Gamma(x_a+1)}$$

*(10)*

To eliminate the nuisance parameter $\lambda$, in the Bayesian paradigm, we just need to integrate with respect to this parameter. We move from a likelihood with 2 parameters $p_{(\Gamma,\Phi)}(\mu,\lambda|x,y)$ to a single-parameter likelihood $p_\Theta(\theta|x,y)$ (a marginal likelihood (Sprott, 2008)).

$$p_\Theta(\theta|x_a,y_a) = \int_0^\infty p_{(\Gamma,\Phi)}(\lambda+\theta,\lambda|x_a,y_a)\,d\lambda$$

$$= \frac{1}{\Gamma(x_a+1)\Gamma(y_a+1)}\int_0^\infty (\lambda+\theta)^{x_a}e^{-(2\lambda+\theta)}\lambda^{y_a}\,d\lambda$$

To our knowledge, there is no simple and general expression for this difference in gamma laws. (Johnson et al., 1994)

### 6.3.1. Hypothesis testing

By analogy with the frequentist case, the hypothesis test that we wish to carry out is to know if $\lambda = \mu$.

We could therefore be interested in the parameter $\theta = \mu - \lambda$ and test the alternative hypothesis ={ $H_1 = \{\theta > 0\}$ }.

This is equivalent to being interested in $\mu > \lambda$ and so $\tau = \frac{\mu}{\lambda} > 1$. The ratio of two gamma laws $Gamma(x_a+1,1)$ et $Gamma(y+1,1)$ is a Beta Prime distribution $\beta'(x+1,y+1)$ whose cumulative distribution function is known (Bourguignon, 2021). The probability that $\tau = \frac{\mu}{\lambda} > 1$ can be expressed as a regularized incomplete beta function:

$$I_{\frac{1}{2}}(x_a+1,y_a+1) = p_{\Gamma-\Phi}(\theta>0) = 1 - p_{\Gamma-\Phi}(\theta<0) = p_{\frac{\Gamma}{\Phi}}(\tau>1)$$

It is therefore possible to define a decision threshold $x_{ac}$, for a given confidence index $\alpha_c$ as :

$$I_{\frac{1}{2}}(x_{ac}+1,y_a+1) = \alpha_c$$

We find the same result for the decision threshold as in the frequentist approach of the previous paragraph with a small difference (the replacement of $x$ by $x_a$ and $y$ by $y_a$). This difference is linked to the choice of the prior (which is logical and natural) and diminishes in influence if $x$ and $y$ are much larger than 1. It is therefore only significant for low counting values. Note also that for a uniform prior ($a=0$) the frequentist and Bayesian approach precisely coincide.

We can express the incomplete regularized beta function as a function of the sum of binomial coefficients (Kirkpatrick & Young, 2009; V. Vivier & Aupiais, 2007):

$$I_{1/2}(x_a+1,y_a+1) = \left(\frac{1}{2}\right)^{x_a+y_a+1}\sum_{i=0}^{x_a} C_i^{x_a+y_a}$$

Where $C_i^{x_a+b}$ is the binomial coefficient:

$$C_i^{x_a+b} = \frac{(x_a + y_a)!}{i!\,(x_a + y_a - i)!}$$

There are methods for inverting the incomplete beta function cumulative distribution function of the beta distribution.(Temme, 1992). In order to obtain the decision threshold, mathematical software also makes it possible to invert this function which is frequently found in the statistical literature.,

Approximations of the beta incomplète regularized function do exist (Abramowitz & Stegun, 1965) :

$$I_x(a+1, b+1) = \Phi\left[3\frac{(bx)^{\frac{1}{3}}\left(1-\frac{1}{9b}\right) - (a(1-x))^{\frac{1}{3}}\left(1-\frac{1}{9a}\right)}{\sqrt{\frac{(a(1-x))^{2/3}}{a} + \frac{(bx)^{2/3}}{b}}}\right] + O(\frac{1}{\min(a,b)})$$

*(11)*

Note that this approximation is considered valid at 0.5% for values of a and b such that a+b>6.

### 6.3.2. Marginal likelihood, binomial expansion and credibility intervals

we will use the binomial expansion of $(\lambda + \theta)^{x_a}$ applied to equation (10)

(Lemay, 2012, 2015; Little, 1982)then marginalize in relation to $\lambda$.

$$p_\Theta(\theta|x_a, y_a) = \int_0^\infty p_{(\Gamma,B)}(\lambda, \mu = \lambda + \theta|x_a, y_a)d\lambda = \frac{1}{\Gamma(x_a+1)\Gamma(y_a+1)}\int_0^\infty (\lambda+\theta)^{x_a} e^{-(2\lambda+\theta)}\lambda^{y_a}\,d\lambda$$

If we restrict ourselves to $x_a \in \mathbb{N}$ et $b \in \mathbb{N}$, $\Gamma(x_a+1) = (x_a)!$ et $\Gamma(y_a+1) = (y_a)!$

We get:

$$p_\Theta(\theta|\,x_a, y_a) = \frac{1}{(x_a)!\,(y_a)!}\int_0^\infty (\lambda+\theta)^{x_a} e^{-(2\lambda+\theta)}\lambda^{y_a}\,d\lambda$$

But

$$(\lambda+\theta)^{x_a} = \sum_{i=0}^{x_a} C_g^i\,\theta^i \lambda^{g-i}$$

Where $C_i^g$ is the binomial. Coefficient.

Therefore:

$$p_\Theta(\theta|x_a, y_a) = \frac{1}{(x_a)!\,(y_a)!}\int_0^\infty \sum_{i=0}^{x_a} \frac{x_a!}{i!\,(x_a-i)!}\theta^i \lambda^{x_a+y_a-i} e^{-(2\lambda+\theta)}\,d\lambda$$

$$= \frac{1}{y_a!}\sum_{i=0}^{x_a} \frac{\theta^i e^{-\theta}}{i!\,(x_a-i)!}\int_0^\infty \lambda^{x_a+y_a-i} e^{-(2\lambda)}\,d\lambda$$

Knowing that $\int_0^\infty x^\nu e^{-\mu x} dx = \frac{\Gamma(\nu)}{\mu^\nu}$, we get for $\theta > 0$

$$p_\Theta(\theta|x_a, y_a) = \frac{1}{y_a!} \sum_{i=0}^{x_a} \frac{\theta^i e^{-\theta}}{i!(x_a-i)!} \frac{(x_a+y_a-i)!}{2^{x_a+y_a-i+1}}$$

*(12)*

We can recognize the product of two terms in this sum. One $\frac{\theta^i e^{-\theta}}{i!!}$, is a Poisson distribution. The other is a negative binomial distribution $\frac{(x_a+y_a-i)!}{y_a! \, 2^{x_a+y_a-i+1}}$

In other words:

$$p_\Theta(\theta|x_a, y_a) = \sum_{i=0}^{x_a} p_S(i|\theta) p_V(x_a-i|y_a)$$

where:

$$p_S(i|\theta) = \frac{\theta^i e^{-\theta}}{i!}$$

Is a term from a Poisson lawAnd

$$p_V(x_a-i|y_a) = \frac{(x_a+y_a-i)!}{y_a!}$$

Is a term from a negative binomial law of parameters $(y_a, \frac{1}{2})$

This is the convolution of a Poisson distribution and a negative binomial distribution. We find again the frequentist expression of the paragraph 6.2.2 taking into account the effect of the chosen priors (which results in the transposition of $x$ to $x_a$ and $y$ to $y_a$). The influence of the priors will only be important for low count values.

We can see that the two methods recommended for getting rid of nuisance parameters (use of conditional likelihood in the frequentist case and marginal likelihood in the Bayesian case) lead to similar results (within the influence of priors).

### 6.3.3. Confidence intervals and detection limits

To obtain the detection limit, it would therefore be necessary, knowing the decision threshold $x_{ac}$, we must find $\theta_d$ such that :

$$\beta_c = \frac{1}{(y_a)!} \sum_{i=0}^{x_{ac}} \frac{1}{i!(x_{ac}-i)!} \frac{(x_{ac}+y_a-i)!}{2^{x_{ac}+y_a-i+1}} \int_{\theta_d}^\infty \theta^i e^{-\theta} d\theta$$

It should be noted that the integral is an incomplete gamma function (Gradshteyn et al., 2000) :

$$\int_{\theta_d}^{\infty} \theta^i e^{-\theta} d\theta = \Gamma(i, \theta_d) = (i)! \, e^{-\theta_d} \sum_{j=0}^{i} \frac{\theta_d^{\,j}}{j!}$$

And so,

$$\beta_c = \frac{1}{(y_a)!} \sum_{i=0}^{x_{ac}} \frac{1}{i!\,(x_a - i)!} \frac{(x_{ac} + y_a - i)!}{2^{x_{ac}+y_a-i+1}} (i)! \, e^{-\theta_d} \sum_{j=0}^{i} \frac{\theta_d^{\,j}}{j!} = e^{-\theta_d} \sum_{i=0}^{x_{ac}} \frac{(x_{ac} + y_a - i)!}{(y_a)!\,(x_{ac} - i)!} \frac{1}{2^{x_{ac}+y_a-i+1}} \sum_{j=0}^{i} \frac{\theta_d^{\,j}}{j!}$$

Knowing the decision threshold $x_{aC}$, the detection limit $\theta_d$ must verify:

$$\beta_c = e^{-\theta_d} \sum_{i=0}^{y_a+k-a} \frac{(x_{ac} + y_a - i)!}{(y_a)!\,(y_a + x_{ac} - i)!} \frac{1}{2^{g_c,+y_a-i+1}} \sum_{j=0}^{i} \frac{\theta_d^{\,j}}{j!}$$

This formula could be evaluated numerically in particular for small values of $y_a$.

As we saw previously, the detection limit is nothing other than the upper limit of the confidence interval for a measurement equal to the decision threshold. To obtain the limits of a confidence interval, it is therefore more generally sufficient to set $\alpha$ and $\beta$ as lower and upper confidence indexes of the confidence interval. It is then necessary to find $\theta_-$ et $\theta_+$, such that for measurements $y$ of the reference and $z + y$ of the sample we obtain

$$\alpha = \sum_{i=0}^{y_a+k-a} \frac{(2y_a + k - i)!}{(y_a)!\,(y_a + k - i)!} \frac{1}{2^{2y_a+k-i+1}} \int_0^{\theta_-} \theta^i e^{-\theta} d\theta$$

And

$$\beta = \sum_{i=0}^{b+k-a} \frac{(2y_a + k - i)!}{(y_a)!\,(y_a + k - i)!} \frac{1}{2^{2y_a+k-i+1}} \int_{\theta_+}^{\infty} \theta^i e^{-\theta} d\theta$$

## 6.4. Synthesis

For low-level metrology with Poisson laws, it is possible to determine a decision threshold from the frequentist point of view and from the Bayesian point of view and then to note their compatibility. Furthermore, we can prove using Neyman Pearson's lemma that this common threshold is the best possible. Likewise, the two approaches (conditional and marginal) lead to very similar results for the elimination of the nuisance parameter and the determination of the likelihood of the desired signal. In all these cases, only measurements with very low counting rates would lead to significant differences in results. It was not possible to define a simple expression detection limit even if it would be possible to determine it numerically.

We can now focus on the transition to the limit of Poisson's laws for large counting values. This will allow us to determine more explicit formulas which will apply for example to the case of measuring radioactivity. It is also possible to determine the distribution of θ based on the measurements $x_a, y_a$:

$$p_\Theta(\theta | x_a, y_a) = \frac{1}{\Gamma(x_a + 1)\Gamma(y_a + 1)} \iint \lambda^{x_a} e^{-\lambda} \mu^{y_a} e^{-\mu} \delta(\theta + \mu - \lambda)\, d\mu d\lambda$$

It is then possible to express this probability distribution as hypergeometric functions (see Annexe 3).

# 7. HETEROSCEDASTIC GAUSSIANS AS A POISSON LAW LIMIT

## 7.1. Current method

### 7.1.1. Decision thresholds and detection limits - the classic Currie approach

The current method is frequentist in its essence and was first developed by Currie (Currie, 1968). The standard in force for measuring radioactivity has colored this approach with considerations that are supposed to be Bayesian. (ISO, 2010a) drawing inspiration from the work of Weise (Weise, 1998; Weise et al., 2006, 2013). We will see what it is in a later paragraph.

In the case of radioactivity measurement, we naturally consider that the desired signal will behave according to Poisson's law. This has an intrinsic uncertainty (its variance is non-zero)

We thus get:

$$p_S(z|\theta) = \frac{\theta^z}{z!} e^{-\theta}$$

For large enough $\theta$, one can use the approximation (Barlow, 1993; Riley et al., 2006):

$$p_S(z|\theta) \sim \frac{e^{-(\frac{(z-\theta)^2}{2\theta})}}{\sqrt{2\pi\theta}}$$

A Poisson distribution for large parameter values θ behaves like a Gaussian with mean and variance θ.

If the baseline also follows a Poisson distribution with a parameter $\lambda$ much larger than 1, we can approximate a Gaussian law:

$$p_B(y|\lambda) = \frac{e^{-(\frac{(y-\lambda)^2}{2\lambda})}}{\sqrt{2\pi\lambda}}$$

Setting:

$$\sigma_B{}^2 = \lambda$$

It is then possible to determine the distribution of the sample as a convolution product:

$$p_G(x|\mu) = (p_{B'} \oplus p_S)(x|\mu)$$

We can determine the convolution of two Gaussians (Bromiley, 2003) which will itself be a Gaussian:

$$p_G(x|\mu) \sim \frac{e^{-(\frac{(x-\mu)^2}{2(\lambda+\theta)})}}{\sqrt{2\pi(\lambda+\theta)}}$$

Where the variance of the sample is the addition of the signal and the reference variances

$$\sigma_G^2 = \lambda + \theta$$
$$\Rightarrow \sigma_G = \sqrt{\lambda + \theta}$$

To obtain the net distribution, we must subtract the noise included in the sample. In terms of distributions as we indicated in the introduction, this corresponds to making a cross-correlation of the distributions, which is itself a Gaussian:

$$p_N(z|\theta) \sim \frac{e^{-(\frac{(z-\theta)^2}{2\sigma_N^2})}}{\sqrt{2\pi}\sigma_N}$$

**(13)**

Where $\sigma_N^2 = \sigma_B^2 + \sigma_G^2 = \sigma_B^2 + \sigma_B^2 + \theta = 2\lambda + \theta$

Indeed, in a difference as in a sum of random variables, the resulting variance is the sum of the variances.

We seek to determine the result of a hypothesis test with the null hypothesis $[\theta = 0]$. It therefore seemed quite natural to consider that (Currie, 1968, 2008) :

$$\sigma_N(\theta = 0) = \sqrt{2\lambda + 0} = \sqrt{2\lambda}$$

Then estimate $\lambda$ by replacing it with the measured value of the reference ($y$, which is the maximum likelihood estimator of the Poisson distribution for the reference). In fact, although this is not always explained in the scientific literature, the different authors find themselves confronted with the problem of an unknown nuisance parameter and assume it to be perfectly known as a solution to the problem (Currie, 1968; Lista, 2016). This is a profile likelihood process (Sprott, 2008) We then fall back on the homoskedastic hypothesis test (equations (3) and (4)) . The formula for the decision threshold is therefore exactly the same as for the homoscedastic case.:

$$z_c = \sigma_N \Phi^{-1}(1 - \alpha_c) = \sqrt{2}\sigma_B \Phi^{-1}(1 - \alpha_c) = \sqrt{2y}\Phi^{-1}(1 - \alpha_c)$$

with $\Phi^{-1}(1 - \alpha_c) = k_{1-\alpha}$

$$z_c = \sigma_N \Phi^{-1}(1 - \alpha_c) = \sqrt{2}\sigma_B \Phi^{-1}(1 - \alpha_c) = \sqrt{2y}k_{1-\alpha}$$

**(14)**

While the detection limit is determined by the equation:

$$\beta_c = \int_{-\infty}^{z_c} p_N(z|\theta_d) dz = \frac{1}{\sqrt{2\pi}\sigma_N}\int_{-\infty}^{z_c} e^{-\frac{(z-\theta_d)^2}{2\sigma_N^2}} dz = 1 - \Phi(\frac{\theta_d - z_c}{\sigma_N})$$

Which leads to the equation:

$$\Phi^{-1}(1 - \beta_c) = \frac{\theta_d - z_c}{\sigma_N}$$

As we have seen, $\sigma_N$ is a function of $\theta$:

$$\Phi^{-1}(1-\beta_c) = \frac{\theta_d - z_c}{\sqrt{2k_{1-\alpha}\sigma_B^2 + \theta_d}} = \frac{\theta_d - z_c}{\sqrt{z_c + \theta_d}}$$

$$(\Phi^{-1}(1-\beta_c))^2 = \frac{(\theta_d - z_c)^2}{z_c + \theta_d}$$

If, as is customary, we take $\beta_c = \alpha_c$

$$\theta_d = 2z_c + (k_{1-\alpha})^2$$

**(15)**

In Currie's approach, these are the formulas we obtain for the decision threshold and detection limit (Currie, 1999b, 2008; Strom & MacLellan, 2001). However, the Gaussian approximation for a Poisson distribution is only valid for large θ (Barlow, 1993; Riley et al., 2006). It is therefore not possible to use it for $\theta \approx 0$ ! CYet this is what is done in the classic Currie approach. Let us repeat that this also assumes perfect knowledge of the reference ($\lambda$), which is in reality very rarely the case.

We also clearly see the conceptual problem; the hypothesis test will in no way depend on the variance of the signal. Whether or not the signal is tainted by significant "noise", the hypothesis test remains the same. In the limit considering zero dispersion for $\theta = 0$ and infinite $\theta > 0$, the hypothesis test would remain the same, while under such conditions it is clear that it would be impossible to differentiate a signal from the baseline.

Several other points remain problematic in the classic approach. They will be addressed in the following paragraphs and we will see how the conditional or marginal likelihood approach provides solutions.

What is done in the standards and reference texts (FDA, 2004; IAEA, 2017) is to consider that the parameter $\theta = \mu - \lambda$ will be expressed as a function $z = x - y$ without further formal derivation ((FDA, 2004; IAEA, 2017; ISO, 2010a) and to assume that the Gaussian approximation of the Poisson law is valid. We therefore obtain the probability density of $z$ in the form of the equation (13) :

$$p_N(z|\theta) \sim \frac{e^{-(\frac{(z-\theta)^2}{2\sigma_N^2})}}{\sqrt{2\pi}\sigma_N} = \frac{e^{-(\frac{(z-\theta)^2}{2(\lambda+\mu)})}}{\sqrt{2\pi}(\lambda+\mu)}$$

The variance is in fact assumed to be proportional to the count values:

$$\sigma_N^2 = \lambda + \mu \sim x + y$$

Which gives, in a Gaussian approximation, confidence intervals for θ of the type:

$$[z \pm k\sigma_N] = [(x-y) \pm k(\sqrt{x+y})]$$

where $k$ is the coverage factor.

As indicated previously, we assume that $\lambda$ is known and can be approximated by $y$ to get rid of the nuisance parameter. Furthermore, we assume that we can also approximate $\mu \sim x$. Note the inconsistency between the confidence interval where a measurement of $z = x - y$ will lead to the possibility of having a confidence interval containing 0 if $z < k\sqrt{x+y}$ and the decision threshold of the previous paragraph which will reject this possibility if $z > k\sqrt{2y}$. This inconsistency is also that of assuming that $\lambda \sim y$ because $\lambda \gg 1$ but not that $\mu = \lambda + \theta \sim x + y$.

### 7.1.2. Bayesian approach in metrology

Some authors conclude that Bayesian statistics encounter too many difficulties to be used in metrology ((Willink, 2010a, 2010b, 2013). The criticisms relate in particular to the fact that the statistical performances are not adequate. As these statistical performances are often frequentist concepts, this type of discussion tends to focus on epistemological principles without it being easy to decide (Bergamaschi et al., 2013; Mana & Palmisano, 2014)

When we consider a parameter which by nature is strictly positive (such as mass or activity), almost unanimous usage dictates that we use a positive support prior.. That is to say, if $\theta$ is the parameter of the desired signal, the prior $\pi(\theta)$ will be defined so that $\pi(\theta) = 0$ if $\theta < 0$. Usually, we use a prior that we will call Heavyside prior defined byr

$$\pi(\theta) = 0 \; si \; \theta < 0$$
$$\pi(\theta) = C \; si \; \theta \geq 0$$

*(16)*

where $C$ is a positive constant. This is the case of the overwhelming majority of publications and standards that use Bayesian methodologies in the field of metrology. (Analytical Method Committee, The Royal Society of Chemistry, 2010; Analytical Methods Committee, The Royal Society of Chemistry, 2008; Bergamaschi et al., 2013; Bochud et al., 2007; Heisel et al., 2009; IAEA, 2017; Kirkpatrick et al., 2013, 2015; Korun et al., 2014, 2016; Laedermann et al., 2005; Lira, 2009; Michel, 2016; Miller et al., 2002; Nosek & Nosková, 2016; Rivals et al., 2012; A. Vivier et al., 2009; Weise et al., 2006; Zähringer & Kirchner, 2008). Some authors have identified difficulties with this prior but attributing these to Bayesian methodology in general. (Willink, 2010b, 2010c, 2013).

This amounts in our formulation presented in paragraph 4.2, to consider that the variable θ must be positive because the desired signal must be positive. Let us insist on the fact that we are taking our desires for realities.. What we want to achieve is the true value, but what we have access to through measurement is an inference on the parameter of the variable $\mathbf{\Theta} = \mathbf{\Gamma} - \mathbf{\Phi}$, the difference between $\mu$ and $\lambda$.

An attempt has nevertheless been made to introduce Bayesian concepts into the definition of characteristic limits (decision thresholds and detection limits): The measurement of ionizing radiation.

### 7.1.3. Pseudo Bayesian approach of the ISO 11929 standard

In the field of radioactivity measurement, the ISO 11929 standard is the reference (ISO, 2010a). It uses the principle of maximum entropy (Jaynes, 2003; Weise et al., 2006) to establish that the probability density is expressed as follows (annex F of (ISO, 2010a)):

$$p_\Omega(\theta|y) = C\pi(\theta)e^{-\frac{(\theta-y)^2}{2u_\theta^2(\theta)}}$$

Where $\pi(\theta)$ is the Heavyside prior : $\pi(\theta) = 0 \; if \; \theta \in ]-\infty, 0]$ and $\pi(\theta) = 1 \; if \; \theta \in [0, +\infty]$. $C$ is a normalization constant and $u_\theta(\theta)$ is the uncertainty for the variable $\theta$.

The principle of maximum entropy determines the type of distribution in the least arbitrary way possible, taking into account all constraints. It actually states that the least arbitrary distribution possible on $[-\infty, +\infty]$ given a mean and a standard error is a gaussian distribution. On the other hand, on $[0, +\infty]$, it is possible to show that the distributions are truncated Gaussian distributions (Dowson & Wragg, 1973). Since the standard deviation is not known since it depends on θ and λ, the authors are forced to assume that $u_\theta(\theta) \sim u_y(y)$ where $u_y(y)$ is the uncertainty of the variable $y$. By normalizing the probability density, we obtain:

$$p_\Delta(\theta|y) = \frac{e^{-\frac{(\theta-y)^2}{2u_y^2(y)}}}{\sqrt{2\pi}\,u_y(y)\,\Phi(\frac{y}{u_y(y)})} \quad \text{avec } \theta \in [0, +\infty]$$

This probability density is used to determine an estimator and a credibility interval. Regarding the estimator ($\hat{\theta} = \hat{y}$ in ISO notations) :

$$\hat{\theta} = \int_0^{+\infty} \theta \frac{e^{-\frac{(\theta-y)^2}{2u_y^2(y)}}}{\sqrt{2\pi}\,u_y(y)\,\Phi(\frac{y}{u_y(y)})} d\theta = y + \frac{u_y(y)}{\sqrt{2\pi}} \frac{e^{-\frac{(y)^2}{2u_y^2(y)}}}{\Phi(\frac{y}{\sqrt{u_y(y)}})}$$

The authors of the standard also calculate a credibility interval $[\theta^\triangleleft, \theta^\triangleright]$.

$$p_\Delta(\theta < \theta^\triangleleft | y) = \int_0^{\theta^\triangleleft} \frac{e^{-\frac{(\theta-y)^2}{2u_y^2(y)}}}{\sqrt{2\pi}u(y)\Phi(\frac{y}{u(y)})} d\theta = \alpha/2$$

$$\Phi\left(\frac{\theta^\triangleleft - y}{u(y)}\right) = \Phi\left(\frac{y}{u(y)}\right)\left(1 - \frac{\alpha}{2}\right)$$

$$\theta^\triangleleft = y - u(y)\Phi^{-1}(\Phi\left(\frac{y}{u(y)}\right)\left(1 - \frac{\alpha}{2}\right))$$

And

$$\theta^\triangleright = y + u(y)\,\Phi^{-1}(1 - \Phi\left(\frac{y}{u(y)}\right)\frac{\alpha}{2})$$

This interval only contains positive terms because, due to the chosen prior, we must have $\theta^\triangleleft > 0 \; \forall y$. The case of credibility intervals for $y < 0$ is evacuated by specifying that this interval must only be calculated for a measurement greater than the decision threshold.

In the ISO 11929 standard, the decision threshold is calculated using a frequentist mode. And for a good reason! Indeed, the choice of the Heavyside prior, by definition prohibits the possibility that the parameter is not greater than zero and therefore that the credibility interval contains 0. The value $\theta = 0$ having a null

Lebesgue measure, no integral equation will be able to tell us that this is possible unless we use not functions but distributions in the sense of Schwarz (Dirac delta function). This is verified in the calculation we have just performed. If we assume that $\theta > 0$, we cannot show that $\theta = 0$ has a non-zero probability. Faced with this impasse, the only solution was to use a frequentist method.

By applying the inverse Bayes' theorem (ISO, 2010a):

$$p_{G-B}(y|\theta) = \frac{e^{-\frac{(\theta-y)^2}{2u_\theta^2(\theta)}}}{\sqrt{2\pi}u(y)}$$

From there, it is possible to calculate the decision threshold as $\alpha_c = p_N(y > y_c|\theta = 0) = \int_{y_c}^{\infty} p_N(y|0)dy = \int_{y_c}^{\infty} \frac{1}{\sqrt{2\pi}u(y)} e^{-\frac{(0-y)^2}{2u_\theta^2(0)}} dy = 1 - \Phi(\frac{y_c}{u(\theta=0)}) \; \theta = 0 \; y_c = u(\theta = 0) \; \Phi^{-1}(1-\alpha_c) = \sqrt{2y} \; \Phi^{-1}(1-\alpha_c)$

As we have seen, this is the application of the frequentist method.

This is not reprehensible in itself and it is inevitable due to the Heavyside prior, but one might wonder the benefit of using a Bayesian formalism to ultimately use frequentist methods. Note that this amounts to making the Currie approximation which leads to the equation (14).

This underestimates the decision threshold and does not give good statistical performances as we will see in numerical experiments.

The detection limit is also determined in a frequentist manner with the same approximation used by Currie.

We can therefore list the errors or problems as follows:

:

- By using the principle of maximum entropy, the standard is limited to the use of truncated Gaussian distributions and only those.
- Furthermore, the probability distribution of maximum entropy is only Gaussian (truncated or not) if the variance and the mean are known
- • It was therefore necessary to assume the known variance, which implies estimating the nuisance parameter (replacing the nuisance parameter with an estimate)
- The prior used is a Heavyside prior which implies that the desired parameter $\theta > 0$. This prior excludes the possibility of using Bayesian methods to determine decision threshold and detection limit.
- The characteristic limits determined are in fact determined by the Currie method. We will see later what are their statistical performances.

## 7.2. Proposed approach using marginal and conditionnal likelihood

We have already seen that in an estimation interval approach, it is not necessary to calculate the decision threshold since the simple view of the confidence or credibility interval is enough to know the significant nature or not of the measurement. .

For comparison with the classical method, we will nevertheless calculate the characteristic limits.

The decision threshold is the smallest measure from which the confidence or credibility interval will include the value zero. As we saw in the paragraph 4.3.1, we want to get a $z_c$ such that for a given confidence index $\alpha_c$, the credibility interval of the net signal barely includes zero:

$$\alpha_c = \int_{-\infty}^{0} p_\Theta(\theta|z_c)d\theta$$

## 7.2.1. Convolution of a Poisson distribution and a negative binomial distribution

We have seen in the case of Poisson distributions that the conditional and marginal likelihoods are convolutions of a negative binomial distribution and a Poisson distribution (Delaporte's law). This convolution has no simple expression. However, the negative binomial distribution converges quite quickly to the normal distribution. This is due to the fact that the negative binomial law $NB(y, p)$ for $p = 1/2$ can be seen as a sum of geometric laws, which, through the central limit theorem, ensures rapid convergence towards a Gaussian distribution(S. Bagui & Mehra, 2019, 2019). (JOHNSON et al., s. d.) gives as a benchmark a value of approximately $y \sim 10$.

$$p_c(y; y) = \frac{(y+y')!}{y!\,y'!}\left(\frac{1}{2}\right)^{y+y'+1} \sim \frac{e^{-\frac{(y'-y)^2}{4y}}}{\sqrt{2\pi 2y}}$$

The convolution product of a negative binomial distribution and a Poisson distribution can then be considered as the convolution product of a Gaussian distribution with mean $y$ and variance $2y$ with a Poisson distribution $S$.

$$p_S(u) = \frac{\theta^u}{u!}e^{-\theta}$$

A Poisson distribution with parameter θ can be approximated by a Gaussian with mean θ and variance θ (Barlow, 1993) only if $\theta$ is large:

$$\frac{\theta^z}{z!}e^{-\theta} \sim \frac{e^{-\frac{(z-\theta)^2}{2\theta}}}{\sqrt{2\pi\theta}}$$

for $\theta \gg 1$

Here, nothing allows us to say that θ is large. Quite the contrary!

On the other hand, if a Gaussian distribution has a sufficiently large variance, we can consider that it can be equivalent to a Poisson distribution. (Figure 6)

Indeed if $y \gg 1$:

$$p_c(y;y) \sim \frac{e^{-\frac{(y'-y)^2}{4y}}}{\sqrt{2\pi 2y}} = \frac{e^{-\frac{((y'+y)-2y)^2}{4y}}}{\sqrt{2\pi 2y}} \sim \frac{(2y)^{y+y'}e^{-2y}}{(y+y')!}$$

We will therefore consider that the Gaussian $p_c$ is the limit of a Poisson distribution of parameter $2y$, for a large $y$.

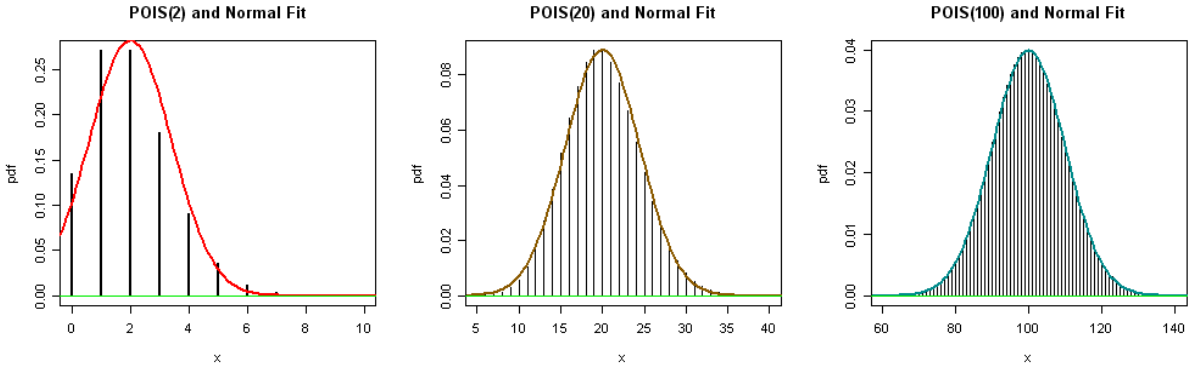

**Figure 6 - Approximation of a Poisson law by a Gaussian for parameter values 2, 20 and 100**

Rather than considering that a Poisson distribution of the signal, poorly measured and of low parameter, can be approximated by a Gaussian distribution, it is more appropriate to consider the Gaussian distribution as the limit of a Poisson distribution. From then on, it becomes easier to make the convolution product of this Gaussian with a Poisson distribution. The convolution of two Poisson laws is itself a Poisson law, with parameter the sum of the parameters (Papoulis, 2002). We therefore obtain the following expression for the conditional likelihood:

$$p_{S+V}(z|\theta) \sim \frac{(\theta + 2y)^{z+2y}}{(z+2y)!} e^{-\theta - 2y}, if\ y \gg 1$$

(17)

This is consistent with a result of (Kruglov, 2012) which characterizes the convolution of a Poisson distribution and a Gaussian as having the characteristic function of a left-shifted Poisson distribution . In a similar way (Koudou & Pommeret, 2002) showed the stability of the Poisson-Gauss family by convolution.

If we use Bayes' theorem with a conjugate prior (Gelman et al., 2013) $\pi(\theta + 2y) = (\theta + 2y)^{u-1} e^{-v(\theta+2y)}$, we then obtain the marginal likelihood of $\theta + 2y$ (pour $v = 0$):

$$p_\Theta(\theta|z, y) = \frac{(\theta + 2y)^{z+2y+u-1}}{(z+2y+u-1)!} e^{-\theta - 2y}$$

(18)

Which is a gamma law of parameters 1 and $z + 2y + u - 1$ for the variable $\theta + 2y$.

And the decision threshold $z_c$ is determined analogously to the equation (5):

$$\alpha = \int_{-2y}^{0} \frac{(\theta + 2y)^{z_c + 2y + u - 1}}{(z_c + 2y + u - 1)!} e^{-\theta - 2y} d\theta$$

Note that the lower bound is $-2y$ and not $-\infty$ because by construction the variable of a gamma law must be positive.

setting $\vartheta = \theta + 2y$, we get

$$\alpha = \int_0^{2y} \frac{(\vartheta)^{z_c+2y+\alpha-1}}{(z_c+2y+u-1)!} e^{-\vartheta} d\vartheta$$

Which can be expressed as a regularized incomplete gamma function:

$$\alpha = \frac{\int_0^{2\sigma^2} (\mu)^{z_c+2y+\alpha-1} e^{-\mu} d\mu}{\Gamma(z_c+2y+u-1)} = \frac{\gamma(z_c+2y+\alpha-1, 2\sigma^2)}{\Gamma(z_c+2y+u-1)} = P(z_c+2y+\alpha-1, 2\sigma^2)$$

Likewise, the detection limit $\theta_d$ will be determined by the equation:

$$\beta = \int_{\theta_d}^{\infty} \frac{(\theta+2y)^{z_c+2y+\alpha-1}}{(z_c+2y+u-1)!} e^{-\theta-2y} d\theta = \int_{\mu_d-2y}^{\infty} \frac{(\mu)^{z_c+2y}}{(z_c+2y+u-1)!} e^{-\mu} d\mu = \frac{\Gamma(z_c+2y+\alpha-1, \mu_d-2y)}{\Gamma(z_c+2y+u-1)}$$
$$= 1 - P(z_c+2y+u-1, \mu_d-2y)$$

From a frequentist point of view, we can from (18) determine a confidence interval in the form (Johnson et al., 1994) :

$$F^{-1}\left(\frac{\alpha}{2}; z+2y, 1\right) < \theta + 2y < F^{-1}\left(1 - \frac{\alpha}{2}; z+2y+1, 1\right)$$

$$F^{-1}\left(\frac{\alpha}{2}; z+2y, 1\right) - 2y < \theta < F^{-1}\left(1 - \frac{\alpha}{2}; z+2y+1, 1\right) - 2y$$

where $F^{-1}$ is the inverse cumulative distribution function of the gamma distribution. Which gives the Bayesian result up to the factor $\alpha - 1$. In particular, there is a coincidence between Bayesian and frequentist results for $\alpha = 1$ (uniform prior) as expected. This result would make it possible to obtain more precise estimation intervals than by considering the Gaussian approximation of the Poisson distribution.

## 7.3. Asymptotic behavior

### 7.3.1. Frequentist asymptotic behavior

From what we have just seen, knowing that $z = x - y$, from the frequentist point of view we have for the equation (18):

$$p_{S+V}(x,y|\theta) \sim \frac{(\theta+2y)^{x+y}}{(x+y)!} e^{-\theta+2y} \sim \frac{e^{-(\frac{1}{2}\frac{(\theta+2y-(x-y)-2y)^2}{(x+y)})}}{\sqrt{2\pi(x+y)}} = \frac{e^{-(\frac{1}{2}\frac{(\theta-(x-y))^2}{(x+y)})}}{\sqrt{2\pi(x+y)}}, \, si \, y \gg 1$$

(19)

While from a Bayesian point of view, the equation (18) becomes

$$p_\Theta(\theta|z,y) = \frac{(\theta+2y)^{x+y+\alpha-1}}{(x+y+\alpha-1)!} e^{-\theta-2y}$$

Which is a shape parameter gamma law $x + y + \alpha - 1$ and scale parameter 1. When $2y \gg 1$, this distribution tends towards a Gaussian (Johnson et al., 1994) with equal mean and variance:

$$p_\Theta(\theta|x,y) \sim \frac{e^{-(\frac{1}{2}\frac{(\theta+2y-(x-y)-2y)^2}{(x+y)})}}{\sqrt{2\pi(x+y)}} \sim \frac{e^{-(\frac{1}{2}\frac{(\theta-(x-y))^2}{(x+y)})}}{\sqrt{2\pi(x+y)}}$$

(20)

### 7.3.2. Confidence intervals

From equation (20), we can identify a pivotal quantity $t = \frac{\theta-(x-y)}{\sqrt{(x+y)}}$ which will have a Gaussian probability density with mean 0 and variance 1:

$$p(t) \sim e^{-\frac{t^2}{2}}$$

Using the properties of this distribution, as in paragraph 3.4, we can set a statistical risk $\gamma$ such that there exists a real $k$ with the constraint:

$$P(-k \leq t \leq k) = \gamma$$

By performing the calculation, we obtain $k = \Phi^{-1}(\gamma)$ where $\Phi^{-1}$ is the inverse of the distribution function of the standard Gaussian distribution (quantile function). $P\left(-k \leq \frac{\theta-(x-y)}{\sqrt{(x+y)}} \leq k\right) = \gamma$

$$P\left(-k\sqrt{(x+y)} \leq \theta - (x-y) \leq k\sqrt{(x+y)}\right) = \gamma$$

$$P\left((x-y) - k\sqrt{(x+y)} \leq \theta \leq (x-y) + k\sqrt{(x+y)}\right) = \gamma$$

For a reference measurement $y$ and a sample measurement $x$, we therefore obtain a confidence interval for $\theta$ : $\left[z - k\sqrt{(x+y)}, z + k\sqrt{(x+y)}\right]$ with coverage probability $\gamma$ and $z = x - y$.

If we compare with the expression recommended by the standards (ISO, 2010b), we see that it is precisely this type of confidence interval that is used for radioactivity measurements. The counting difference $(x - y)$ is used as an estimator of $\theta$ and $\sqrt{(x+y)}$ is used as the uncertainty of this diffErence.

### 7.3.3. Decision threshold obtained from the confidence interval

To determine the decision threshold, we need to determine what would be the smallest value of $z$ (or equivalently, $x$ because $y$ is known) authorizing $\theta = 0$ to belong to the confidence interval.

This corresponds to the lower bound of the confidence interval being zero:

$$x_c - y - k\sqrt{(x_c + y)} = 0$$

thus :

$$z_c = k\sqrt{(z_c + 2y)}$$

(21)

And so:

$$z_c^2 = k_{1-\alpha_c}^2 (2y + z_c)$$

with $k_{1-\alpha_c} = \Phi^{-1}(1-\alpha_c)$.

The solution to this quadratic equation is:

$$z_c = \frac{k_{1-\alpha_c}^2 + \sqrt{k_{1-\alpha_c}^4 + 8k_\alpha^2 y}}{2}$$

*(22)*

We find the expression of equation (21) by different authors (Altshuler & Pasternack, 1963; Alvarez, 2007; Turner, 2007). These authors take heteroscedasticity into account in a somewhat ad hoc manner by postulating that the uncertainty of $p_N(z|\theta)$, the probability density of $z$, is $\sigma_N = \sqrt{2\sigma_B^2 + z}$. They start from the expression $\sigma_N^2 = 2\sigma_B^2 + \theta$ where $\sigma_B$ is the uncertainty of the reference and do the only thing possible without knowing $\theta$ : approximate it by $z$ (the parameter of the $z$ distribution is replaced by the measurement $z$).

### 7.3.4. Decision threshold resulting from the conditional likelihood of Poisson laws

We shave seen (equation (9) from paragraph 6.2) that the decision threshold can be obtained from the conditional likelihood in the case of Poisson distributions in the form :

$$I_{\frac{1}{2}}(x_c + 1, y + 1) = \alpha_c$$

Where $I_{1/2}(x_c + 1, y + 1)$ is the regularized incomplete beta function, $x_c$ is the decision threshold for $x$ and $\alpha_c$ is the confidence index of the hypothesis test. Furthermore, we saw that it was possible to obtain an approximation of the regularized incomplete beta function (equation(11))

Using this approximation, and using taylor expansions in $\frac{z_c}{y}$, we get:

$$1 - \alpha_c \sim \Phi\left(\frac{z_c}{\sqrt{2y}}\right)$$

thus :

$$z_c = \sqrt{2y}\,\Phi^{-1}(1 - \alpha_c)$$

We therefore find in this approximation for large values of x and y, equation (6) (homoscedastic) if $(\sigma_G^2 + \sigma_B^2) \sim 2(\sigma_B^2) = 2y$. This corresponds to Currie decision threshold (14) which is therefore only a first order approximation in $\frac{z_c}{y}$.

If we continue the expansion to the next order we find:

$$1 - \alpha_c \sim \Phi\left(\frac{z_c}{\sqrt{2y + z_c}}\right)$$

And:

$$z_c = \sqrt{2y + z_c}\,\Phi^{-1}(1 - \alpha_c)$$

**(23)**

We find the expression (21), knowing that $z_c = x_c - y$.

We can note that the decision threshold thus determined indicates that the Currie-11929 decision threshold underestimates false positives and will therefore declare results that are not significant to be significant. In fact, this Currie decision threshold is lower than the one we determined and will therefore consider more measurements significant.

### 7.3.5. Detection limits

We know that under present conditions:

$$\theta_d = 2z_c = k_{1-\alpha_c}^{\,2} + \sqrt{k_{1-\alpha_c}^{\,4} + 8k_{1-\alpha_c}^{\,2}\,y}$$

**(24)**

Indeed, the confidence and credibility intervals of a Gaussian distribution used here are symmetrical. So, an interval with zero lower bound, of estimator $z = z_c$ will have an upper bound $2z_c$.

### 7.3.6. Optimal Test

The approach presented in the previous paragraph amounts to considering a Gaussian probability density with an expectation $\lambda$ and a variance $y$ as well as an alternative probability density of expectation $\mu$ and variance $y + z$ as a first approximation. The Neyman-Pearson lemma (see annex 4) insures that the best test of the hypothesis $\lambda = \mu$ is the z-test of different variances (Moore et al., 2009). This test will give (21) ensuring that it is the best hypothesis test. This will therefore be valid to the extent that the second order approximation in, $z/\sigma$ is valid.

This ensures that the decision thresholds and detection limits previously determined by (23) and (24) are optimal.

This also allows us to say that the best test to know whether we should reject a hypothesis of the type $\theta = \theta_o$ will be the inclusion of $\theta_o$ in the confidence interval calculated from paragraph 7.3.2. No other test will perform better. Indeed, this confidence interval is based on the probability density $p_W(x|y,\theta)$.. The best test to know if a result is significant is therefore to check that 0 is not included in the confidence interval. No other test can have better performance. In particular, normative decision thresholds will have worse statistical performances.

### 7.3.7. Bayesian asymptotic behavior

For large $k$, the gamma functions $f(\lambda|k) = \frac{\lambda^k}{k!}e^{-\lambda}$ tend towards Gaussians of average $k$ and variance $k$ (S. C. Bagui & Mehra, 2016; Barlow, 1993). So, for large $x$ and $y$:

$$p_\Gamma(\mu|x) = \frac{\mu^x e^{-\mu}}{x!} \sim \frac{1}{\sqrt{2\pi(x)}} e^{-\frac{(\mu-x)^2}{2(x)}} \quad et \quad p_\Phi(\lambda|y) = \frac{\lambda^y e^{-\mu}}{y!} \sim \frac{1}{\sqrt{2\pi(y)}} e^{-\frac{(\lambda-y)^2}{2(y)}}$$

By calculating the cross correlation, we therefore obtain the formula:

$$p_\Theta(\theta|x,y) = \frac{1}{\sqrt{2\pi(x+y)}} e^{-\frac{(\theta-x+y)^2}{2(x+y)}}$$

Lemay had already observed that this formula was a very good approximation of $p_\Theta(\theta|x,y)$ (Lemay, 2012, 2015).

Using equation(3) to determine the decision threshold, it is therefore necessary to find $z_c$ with $x = y + z_c$ such that

$$z_c = \sqrt{(2y + z_c)}k_{\alpha_c}$$

**(25)**

If we note, as it is traditionally done:

$$k_{\alpha_c} = \Phi^{-1}(1 - \alpha_c)$$

Then, $z_c^2 = k_{\alpha_c}(2y + z_c)$.

By solving the quadratic equation, we have:

$$z_c = \frac{k_{\alpha_c}^2 + \sqrt{k_{\alpha_c}^4 + 8k_{\alpha_c}^2 y}}{2}$$

**(26)**

This expression is equivalent to those of (Altshuler & Pasternack, 1963; Alvarez, 2007; Turner, 2007) with one essential difference already mentioned. The decision threshold here only depends on observation $y$ while, in the mentioned references, it depends on an unknown parameter value which is approximated by observations.

The detection limit expression is given by:

$$\theta_d = z_c + \sqrt{(2y + z_c)}\Phi^{-1}(1 - \beta_c)$$

**(27)**

If we chose $\beta_c = \alpha_c$, we get:

$$\theta_d = k_{\alpha_c}^2 + \sqrt{k_{\alpha_c}^4 + 8k_{1-\alpha_c}^2 y} = 2\, z_c$$

<div align="center">(28)</div>

While traditional approaches are forced to use formulas including unknown parameters that can only be estimated, this is not the case here. We are therefore in the presence of exact formulas, for large $y$.

It is now useful to see how these proposed characteristic limits differ from established limits (Currie-11929) from the point of view of the consequences but also of their statistical performance or in practice.

## 7.4. Detection limit divergence

Several authors have noted that, under certain conditions, the detection limit could diverge (Kirkpatrick et al., 2013, 2015).

It is common in usual metrological situations to assume that the desired parameter depends on the quantity measured up to a multiplicative calibration variable. If, for example, counts are carried out, the final parameter sought is the activity $a$ which will be expressed according to the number of net counts measured $z$:

$$av = z$$

where $v$ is a calibration coefficient.

It is then possible to determine the detection limit in terms of activity and no longer counting. However, the uncertainty in the calibration coefficient must be taken into account.

This is commonly done by adding to the uncertainty (Dietrich, 1991; Kirkpatrick et al., 2013) a term of the form $\sigma_v^2 a^2$. Determination of the detection limit in terms of activity $a_d$ is then done in the form of the solution of the quadratic equation (following the Currie approach used to determine the equation (4):

$$\theta_d = a_d v = z_c + \sqrt{(\sigma_o^2 + a_d v + \sigma_v^2 a_d^2) k_{1-\beta_c}}$$

where $z_c = k_{\alpha_c} \sigma_0 = k_{\alpha_c} \sqrt{2b}$

We then get the solution

$$a_d = \frac{\frac{2\, k_{\alpha_c}\sigma_0 + k_{\beta_c}^2}{v} + \sqrt{(\frac{2\, k_{\alpha_c}\sigma_0 + k_{\beta_c}^2}{v})^2 - 4(1 - k_{\beta_c}^2 \frac{\sigma_v^2}{v^2})(k_{\beta_c}^2 - k_{\beta_c}^2) \frac{\sigma_o^2}{v^2}}}{2(1 - k_{\beta_c}^2 \frac{\sigma_v^2}{v^2})}$$

The problem appears because the term $(1 - k_{\beta_c}^2 \frac{\sigma_v^2}{v^2})$ in the denominator can be zero when $1/k_{\beta_c}^2 = \frac{\sigma_v^2}{v^2}$, causing the detection limit expressed in terms of activity to diverge. The decision threshold can then be calculated ($a_c = \frac{k_{\alpha_c}\sigma_0}{v}$) but not the detection limit. This situation is considered non-physical and attempts have been made to overcome this difficulty (Kirkpatrick et al., 2015). In fact, the decision threshold calculated in this way does not take into account the uncertainty of the calibration coefficient. If this is taken into account, we can determine a decision threshold on the activities.

The uncertainty of the activity $\sigma_a$ can be expressed using relative uncertainties (Dietrich, 1991):

$$\frac{\sigma_a^2}{a^2} = \frac{\sigma_v^2}{v^2} + \frac{2\sigma_N^2}{z^2}$$

The decision threshold on the activity is then such that:

$$a_c = k_{\alpha_c} \sigma_a$$

And thus:

$$\frac{1}{k_{\alpha_c}^2} = \frac{\sigma_v^2}{v^2} + \frac{2\sigma_N^2}{z^2}$$

**(29)**

We can see that when $\frac{1}{k_{\alpha_c}^2} = \frac{\sigma_v^2}{v^2}$ no value of z will be able to verify the previous equation. In this case the relative uncertainty is greater than 100%! No measurement of the activity will ever be significant. The decision threshold diverges and it is therefore perfectly logical that the same applies to the detection limit. No theoretical value will be able to give a significant measured activity. Under such conditions of uncertainty, all measurements are insignificant because we can never be certain that they stand out from the background noise. On the other hand, for each measurement it is possible to give a confidence interval and therefore an upper limit.

## 7.5. Decision thresholds comparison

Figure 7 illustrates the differences between the decision thresholds calculated by the formula (14) due to Currie and that given by the interval method (26) which coincides with Althulser's formula (Altshuler & Pasternack, 1963; Strom & MacLellan, 2001)). They were calculated for $k_{1-\alpha_c} = 2$. We observe that the Currie decision threshold tends towards zero for $y$ tending towards zero. Currie's calculation is based on the fact that $y$ is large and is therefore no longer valid within this limit. It is therefore not surprising that it predicts that the decision threshold disappears for $y = 0$. The Currie decision threshold is also always smaller than the threshold from the intervals. It is underestimated.

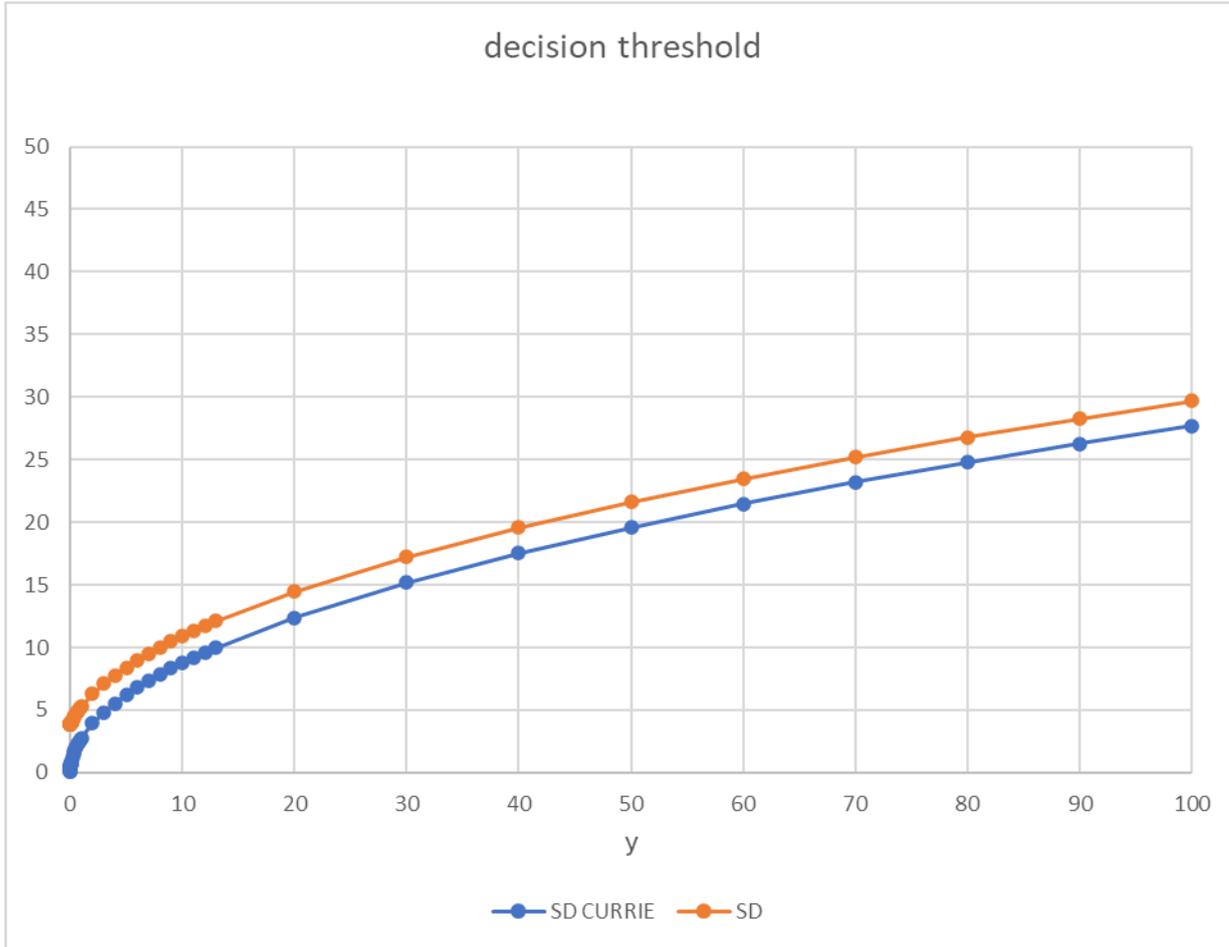

**Figure 7 - Decision thresholds calculated according to Currie (blue curve) and according to the proposed method (red curve)**

Figure 8 illustrates the detection limits calculated using the different methods. It appears that the detection limits according to Currie is very slightly underestimated. Remember that quantitative differences are not the main thing. The important thing is that all the necessary information is contained within the confidence interval. There is no need to calculate a decision threshold or detection limit. The non-significant nature of a measurement is determined by the inclusion of zero in the confidence interval and this is the optimal way to do it. While the upper limit of the interval will give us a smaller upper bound for the estimation of the desired parameter $\theta$.

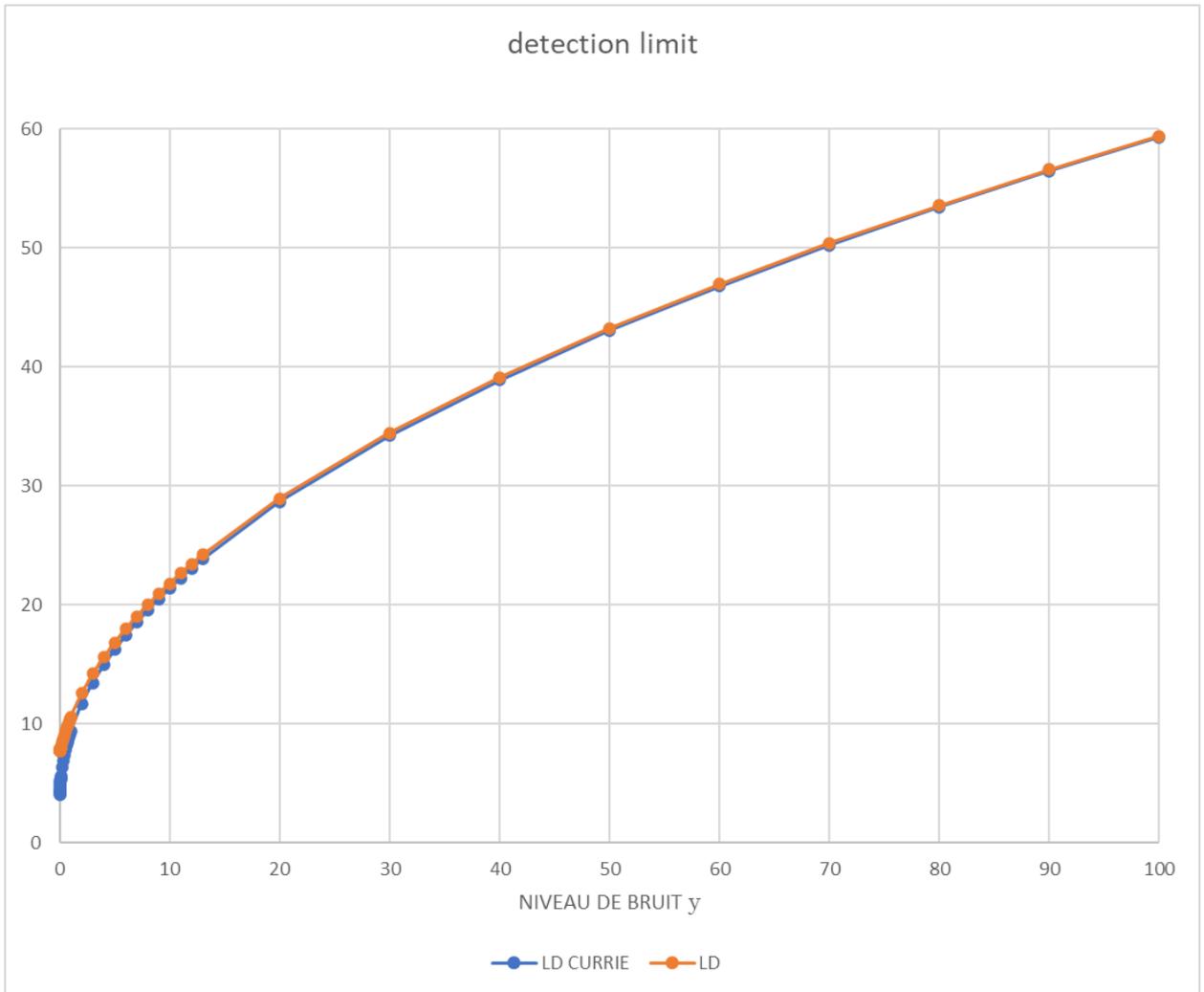

**Figure 8 - Detection limits calculated according to Currie (blue curve) and the proposed method (red curve)**

## 7.6. Influence of the prior

The commonly used priors for a Poisson distribution are the uniform prior, the Jeffreys prior and the inverse prior corresponding respectively to $a = 0$, $a = 1/2$ and $a = 1$ fr a prior $\pi(\lambda, \mu) = \frac{1}{(\lambda\mu)^a}$. The calculations carried out show that for these priors, their influence on the decision thresholds and detection limits consists of adding $a$ to count values $(y, x)$. This has an influence on low count values but very quickly this influence becomes negligible as $y$ increases.

The choice of the prior is therefore not decisive apart from low count values. This is consistent with what is encountered for a simple measurement of a single Poisson distribution (Bolstad, 2007).

Note that it would be possible to take into account possible different experiment times for the measurement of the reference and the sample. To do this, it is enough to extrapolate the counts proportionally to the same measurement of time. If we measure $x$ et $y$ during $t_G$ and $t_B$, then simply choose a reference time and calculate the counts for this value.

## 7.7. Coverage probability

Willink studied the probability of recovery of Bayesian credibility intervals obtained with a Heavyside prior for Gaussian or Student distributions. (Willink, 2010b). Recall that by construction the frequentist confidence intervals for a probability $\alpha$, if the experiments were repeated multiple times, should include the true value $100(1-\alpha)\%$ on average. Through Monte Carlo simulations, Willink observed that this is not the case for credibility intervals when the sample measurement approaches that of the reference. He even showed that when the difference tends towards zero, the credibility intervals have an overlap probability which tends towards zero! These intervals therefore no longer have any chance of containing the true value in the frequentist sense of the term. As with his other observations, Willink attributed this inconsistency to Bayesian methodologies in general (particularly those employing so-called "objective" priors). Other authors have been interested in the question. Some explain this inconsistency by the fact that the confidence and credibility intervals do not answer quite the same question. (Mana & Palmisano, 2014). They were interested in Gaussian distributions. Others, conversely, draw inspiration from methods developed in particle physics in a frequentist framework to modify the credibility intervals and thus ensure a satisfactory coverage probability. (Lee et al., 2016). The elements we provided on the Heavyside prior explain the problem.

In fact, we have seen for the case of Gaussians, that the confidence and credibility intervals coincide if we use a uniform prior (including negative values). By construction, the coverage probability of confidence intervals will be $100(1-\alpha)\%$. Credibility intervals being identical to confidence intervals in this case, we are therefore certain, without even having to carry out simulations, that the coverage probability of these credibility intervals is correct! The inconsistency is therefore lifted for the case of Gaussians. We did not perform the calculation for Student distributions, but the credibility and confidence intervals of Student distributions coincide for a uniform prior. It is when a Heavyside prior is used that coverage inconsistencies appear for the credibility intervals. The conclusion is therefore identical: the Heavyside prior is the source of the problem.

Coverage probabilities are therefore perfectly adequate for the credibility intervals of the Gaussian and Student distributions if we reject the Heavyside priors and use priors including a negative part.

It is the case for the credibility intervals of the random variable Θ and the confidence intervals of the random variable $N$. On the other hand, if we consider Poisson laws, this is not the case. Their supports must be positive. They will be included in the support materials Θ and $N$. This implies that the respective intervals will have a greater coverage If $[\alpha, \beta]$ is the 95% credibility interval for Θ with $\alpha < 0$, we can say that the positive values will be in an interval of $[0, \beta]$ but without being able to guarantee that exactly 95% values will be included in this interval. On the other hand, at least 95% of the values will be there. We say that the confidence interval « overcovers ». In the absence of data on the contribution of noise to the sample measurement, it is not possible to deconvolve Γ ou Θ

## 7.8. Numerical Validation

It is possible to carry out simulation tests to validate the results obtained. Here we adopt Strom's approach (Strom & MacLellan, 2001). This involves starting from a given background noise distribution and determining the decision threshold after drawing from this distribution. Secondly, a new draw is carried out and this draw is compared to the decision threshold previously determined with the same distribution. We are therefore in the presence of background noise. Logically, a result deemed to be significant is therefore a false positive. The use of decision thresholds should therefore give a false positive rate equal to $\alpha$, the chosen confidence index.

For homoscedastic distributions, this is indeed the case (Voigtman, 2017). If we are interested in heteroskedastic distributions (which is the case for radioactivity), it is possible to consider Poisson distributions. First of all, a pure radioactive source can be considered as having an activity following a Poisson distribution. On the other hand, for high activities, its distribution can be approximated by a Poisson Gauss distribution (Gaussian distribution with a heteroscedastic standard deviation). We therefore tested the Currie decision thresholds (which are those of ISO 11929) and those determined by our method. We will limit ourselves here to testing the Poisson-Gaussian approximations of the decision thresholds. We will determine the ratio of the false positive rate between the observed value and the theoretical value for a given value of the background noise parameter. This rate should be as close as possible to 1. We have made a million draws of the background noise. For each draw, we determine the decision threshold, then we make a new draw which we compare to the previously calculated decision threshold.

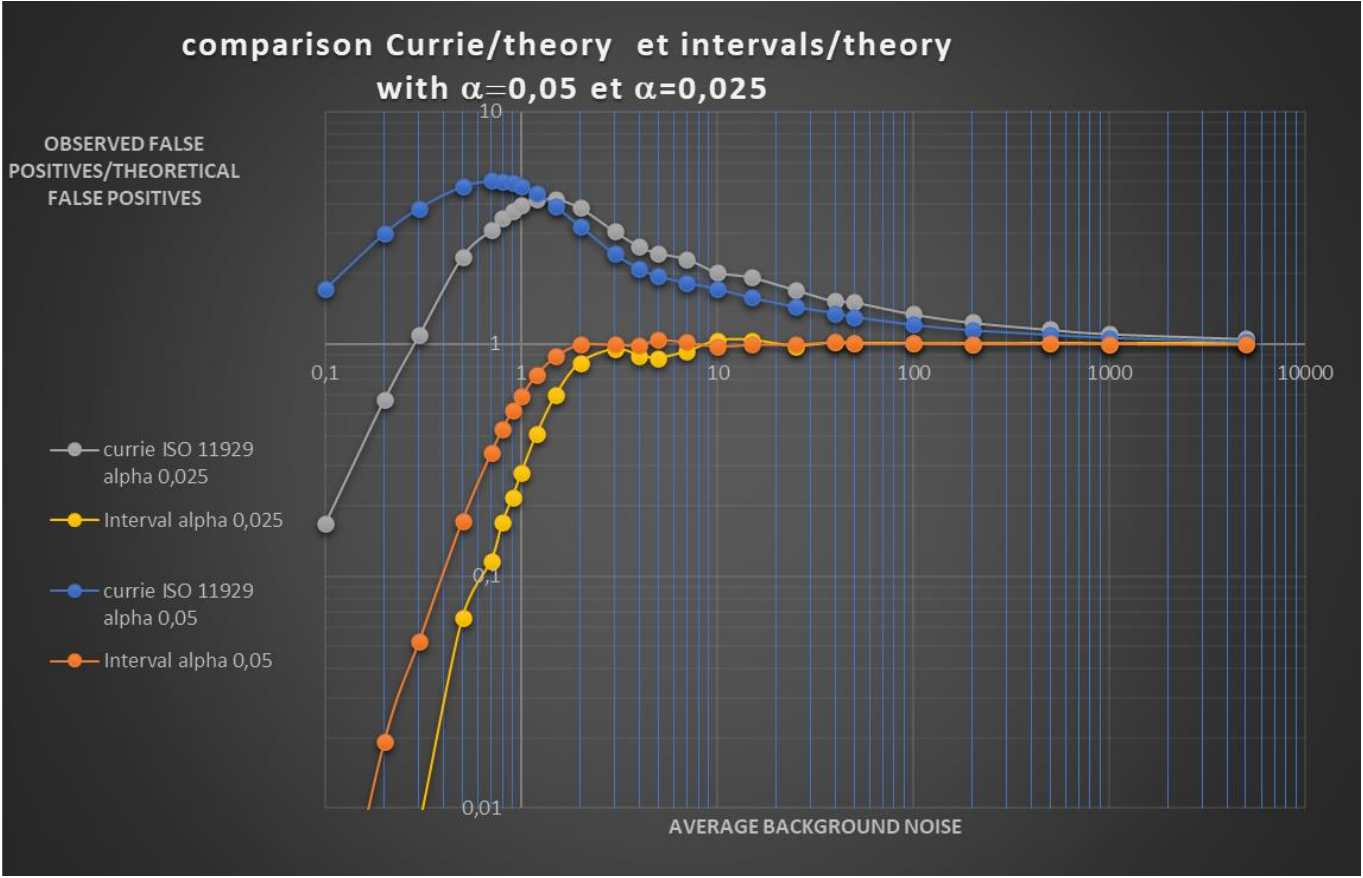

**Figure 9 - Comparison of the ratio of real/theoretical false positives**

We observe that Currie decision thresholds are only effective from very high counts of background noise (larger than 1000). For such values the influence of heteroscedasticity becomes weak compared to the value of the background noise. The distribution becomes almost homoscedastic. It is therefore logical that the Currie thresholds determined with a homoscedastic hypothesis become more efficient. The thresholds derived from the confidence intervals are just as efficient for these high values of the background noise. Below these values, we notice, as expected, that the false positive rate observed in the case of Currie is significantly higher than the theoretical rate α. This corresponds to an underestimation of the decision threshold since there are more measurements from the background noise which are considered significant. This therefore confirms the calculations carried out previously, in particular the fact that the Currie decision threshold is an underestimate of the decision threshold.

This is not the case for the decision thresholds that we determined from the confidence intervals which remain efficient up to low values of the background noise. It is interesting to note that the false positive rate remains consistent with the theoretical value up to values of the background noise in the domain of validity of the Stirling approximation and the approximation of the equation (11). For very low background noise values, the observed background noise counting statistic becomes low and the decision thresholds are impacted. Indeed, it becomes more and more likely that the observed background noise is zero. At this point, for the Currie formula, the decision threshold is zero (any non-zero measurement becomes significant). This leads to an excess number of false positives (up to 25 times the theoretical value). In equation (26) terms including $k_{1-\alpha_c}$ become predominant when the background noise is low and thus lead to decision thresholds higher than those of Currie guaranteeing that the decision thresholds are not too low for zero count. However, for very low background noise values, this threshold becomes too high. It would be possible to use the results of paragraph 6 to determine decision thresholds valid for low background noise values. The equations would need to be solved numerically for each background count value. We didn't do it here. We can therefore confidently consider that the statistical performance of the decision thresholds calculated with our method is much better than that of Currie and ISO 11929, which validates the approach.

This has implications for biological dosimetry. Indeed, for the determination of decision thresholds in the measurement of chromosomal aberrations, it is customary to consider possibly expanded Poisson distributions. We measure the number of chromosomal aberrations in an unexposed population and compare it to that of an exposed population. The simulation carried out therefore makes it possible to establish that up to low average values (~3) of chromosomal aberrations in the unexposed population, the decision thresholds calculated using Bayesian methods with Gaussian Poisson approximation are efficient.

## 7.9. Experimental validation

IRSN ( Institut de radioprotection et de sureté nucléaire) carries out a very large number of radioactivity measurements in the environment with very low activities.

For example, in 2019 an environmental sample (S19EEA21-98B1) was measured in the laboratory using liquid scintillation. The goal was to determine the activity of the tritium present in it. The decision threshold was determined according to the ISO 11929 standard as being worth 0.289 counts per minute (cpm) for the test portion. Measurement by liquid scintillation gives 0.300 cpm. It was therefore concluded in accordance with existing standards that the activity of the sample was significant. The confidence interval was therefore always calculated according to existing standards. It is 0.300+/-0.305 cpm or [-0.005; 0.605]. 0 is included in this confidence interval. The result is therefore not significant since zero is a value compatible with the measurements. We cannot reject the null hypothesis (no activity in the sample). The relative uncertainty is 102%. In fact, there is conflict and inconsistency between Currie's hypothesis test and the confidence interval which, let us remember, is nothing other than the set of values of the parameter which verifies a hypothesis test with complete uncertainty. The measuring laboratory, considering that this result was not presentable, decided, as is usual in these cases, to remeasure the sample. This second measurement confirmed that the activity was not significant. This inconsistency between confidence interval and decision threshold was highlighted in the paragraph 7.1. We therefore note experimentally that indeed the decision thresholds calculated according to standard 11929 (and which are those of Currie) are underestimated and give false positives that the confidence interval criterion would consider to be insignificant. In addition, it is clear that measurement laboratories then take into consideration the confidence interval criterion by rejecting results whose uncertainty is greater than 100%. It would therefore be much simpler and faster to simply calculate the confidence interval directly. It makes it possible to determine the significant nature or not of the measurement, gives an estimate of the activity, avoids making a second measurement and can be used subsequently whether it is significant or not.

**0 included in the confidence interval = uncertainty > 100% = non significant sample.**

The sample result sheet is attached to this document

## 7.10.   . Conclusions on the heteroskedastic case

Let's summarize:

- Traditionally used hypothesis tests (Currie-ISO 11929) do not have the expected statistical performance.
- The use of credibility or confidence intervals, formed from conditional or marginal likelihood, makes it possible to extract the characteristic limits (decision thresholds and detection limits) but this time with correct statistical performance.
- Neyman Pearson's lemma ensures that this is the best possible test.
- The simple observation of the estimation interval (Bayesian or frequentist since they almost coincide except for very low counting values) makes it possible in any case to determine the significant nature or not of the measurement.
- These same intervals make it possible to give an upper limit to our parameter which depends on the measurement and which is more precise than a detection limit.
- In practice, the use of Currie's decision threshold, which is in fact optimal only for the homoskedastic case, can lead to experimental inconsistencies in a heteroscedastic case. These are resolved in practice by remeasurements. This inconsistency obviously does not exist when we use confidence intervals as hypothesis tests or, equivalently, when the intervals are based on hypothesis tests. To the extent that the proposed test coincides with that resulting from the Neyman-Pearson lemma, it must logically have better statistical performances. Experience therefore confirms this conclusion.
- Theory, experiment and simulations therefore converge towards the fact that the hypothesis test proposed here is the best.
- Cases where the detection limit diverges. exist The use of conditional or marginal likelihoods makes it possible to eliminate the nuisance parameters, to explain the origin of this difficulty and to resolve it while always making it possible to give an upper limit to the parameter.

# 8.   DATA RENDERING

## 8.1. Current situation

Currently, in metrology, a measurement result $z$ not statistically significant ($z < z_c$) is returned in the form $< \theta_d$ (where $\theta_d$ is the detection limit). In mathematics, we call this type of result censored data.. To be precise, this is left-censored data. The logic behind is to consider that if a measurement is lower than the decision threshold, then only parameters larger than $\theta_d$ will have a probability of at least $100.\beta_c\%$ to give this measurement. The difficulties of dealing with mixtures of significant and non-significant data have been widely discussed elsewhere(Helsel & Helsel, 2012). Their principle is to try to reconstruct the missing data based on hypotheses about the distribution of the underlying parameter. The performance of these methods strongly depends on the percentage of censored data and even the type of methods depends on this percentage. This makes automating data processing extremely difficult. Indeed, the simple addition of a single result can modify the processing method. There is an extensive literature on processing censored data and a large number of methods. Let us repeat that they were originally developed as a stopgap when it was impossible to have uncensored data. In particular, the most used method - substitution - is the one which is unanimously considered the worst by statisticians (Helsel, 2006; Helsel & Helsel, 2012). It consists of a censored result $< \theta_d$ to replace it with $\theta_d$. Any statistician questioned on the subject would be stunned to learn that measurement data is thus censored and not used. He would retort that, in all cases, it is preferable to use basic data to carry out statistical processing.

The first remark that can be made is that the hypothesis test in no way invalidates the measurement result. This one is what it is. The measurement results remain perfectly usable for data processing (average, spatial trend, temporal trend, limit, etc.). It is completely rigorous and scientifically accurate to return a measurement result below the decision threshold. An uncertainty must always be given with a measurement result, including this one.

This is also what is done in the field of particle physics. The established practice is to always return the measurement results independently of the inference made from them. (Anselmann et al., 1995; James & Roos, 1991). This even if the measurement result is negative. This is how, for example, the squared experimental mass of the neutrino is currently recorded in the community of particle physicists as being $-0,6 \pm 1,9\ eV^2$ (Beringer et al., 2012). The fact that a squared mass cannot be negative is not unknown to particle physicists… Simply, the confidence interval covers the zero value and therefore experimentally, taking into account the uncertainties, the zero value is compatible with the experimental results.

We can find several recommendations not to censor the data. The ironic thing is that the very creator of the detection limits and decision thresholds (Lloyd Currie) also strongly recommended returning the results without censoring them! It is worth mentioning this because some are convinced that uncensored restitution is incompatible with the concepts of characteristic limits (decision thresholds and detection limits). This is explained without any ambiguity by Currie in an article (Currie, 2008) on nuclear measurements for radioactivity in the environment.

We reproduce below the main passage on this subject:

### 8.2.1. *Statement of the problem; values and non-values*

Quantifying measurement uncertainty for low-level results—i.e., those that are close to detection limits—deserves very careful attention: (a) because of the impact of the blank and its variability, and (b) because of the tendency of some to report such data simply as "non-detects" (Lambert et al., 1991) or "zeroes" or "less than" (upper limits). The recommendations of IUPAC (1995, 1998) in such cases are unambiguous: experimental results should not be censored, and they should *always* include quantitative estimates of uncertainty, following the guidelines of ISO-GUM (ISO, 1995). When a result is indistinguishable from the blank, based on comparison of the result ($\hat{L}$) with the critical level ($L_C$), then it is important also to indicate that fact, perhaps with an asterisk or "ND" for not detected. But "ND" should never be used alone. Otherwise there will be information loss, and possibly bias if "ND" is interpreted as "zero".

Many scientific organizations recommend reporting measurement results in full with their uncertainty. First of all the IUPAC-International Union of Pure and Applied Chemistry- in its "Orange book" which brings together all its recommendations in terms of chemical analysis (Inczedy et al., 1998). Here again, it is useful to reproduce the passage from 18.4.3.7 devoted to the restitution of non-significant results:

> (2) A result falling below $L_C$, triggering the decision "not detected" should not be construed as demonstrating analyte absence. (See section 18.4.3.6.) Reporting such a result as "zero" or as "<$L_D$" is *not* recommended; the estimated value (net signal, concentration) and its uncertainty should *always* be reported.

In a similar way, but less explicitly, the ISO guide allows no exception in its recommendation to always give a measurement result with its uncertainty. (JCGM, 2008). For radioactivity measurements, as mentioned, the ISO 11929 standard requires the determination of the confidence interval in all cases. On the other hand, it is not very explicit on the restitution of data, simply mentioning that the customer must be provided with any requested information.

Despite all this and in a scientifically incomprehensible way, this recommendation (or even this normative requirement) is not generally followed in Europe.

This is not the case in the USA in the field of radioactivity. There is an interagency document in the USA, the MARLAP-Multi-Agency Radiological Laboratory Analytical Protocols Manual (FDA, 2004). This applies to the DOE (Department of Energy), the USGS (United States Geological Survey), the DOD (Department of Defense), the FDA (Food and Drug Administration), the NRC (Nuclear Regulatory Commission) , to the EPA (Environmental Protection Agency) and to NIST (National Institute of Standards and Technology equivalent of LNE), in short to any federal scientific institution that measures radioactivity in the USA. Any laboratory or organization carrying out field or laboratory measurements on behalf of one of these government organizations should comply with this principle of full restitution. ((FDA, 2004) paragraph 19.3.9 recommendations) :

- The laboratory should report all results, whether positive, negative, or zero, as obtained, together with their uncertainties.

Additional guidance is added in paragraph 20.3 (where MDC is the notation adopted for the detection limit): *The laboratory should report each measurement result and its uncertainty as obtained (as recommended in Chapter 19) even if the result is less than zero. The laboratory should <u>never</u> report a result as "less than MDC"*

It cannot be any clearer.

## 8.2. Current normative requirements for data restitution in the field of radioactivity

Even if the scientific basis of the ISO 11929 standard is incorrect, does it accept the fact of restoring the measurement results in full?

Based on the current standard (ISO 11929 :2010 (ISO, 2010a)), it can be observed that it is mentioned in clause 7:"*The content of the test report depends on the specific application as well as on demands of the customer or regulator*"

As long as there is agreement from the client (which is easy for an internal client), it is therefore possible to mention the desired information. In the same clause 7, it is written that one must always determine and keep in a document:

"d) the primary measurement result, y, and the standard uncertainty, u (y) , associated with y"

So determining the measurement and its uncertainty is in any case a normative requirement. If we continue reading this clause, we find that we must also mention:

"*h) a statement as to whether or not the physical effect is recognized as being present;
NOTE If the physical effect is not recognized as being present, i.e. if y< y\* [ce qui correspond au seuil de decision $z_c$] , it is occasionally demanded by the regulator to document < y# [correspondant à la limite de détection $\theta_d$] instead of the measured result, y. Such documentation can be meaningful since it allows, by comparison with the guideline value, to demonstrate that the measurement procedure is suitable for the intended measurement purpose.*"

This means that the only thing that is required at the level of the standard for a non-significant result in addition to the results and the uncertainty, is a mention of this non-significant character. If the authority requires it, it is possible to return in the form $< \theta_d$ but this is just an additional option. There is therefore no normative obstacle to returning the results in an uncensored manner. It is even rather the opposite, the censored restitution is a tolerance compared to what should ideally be restored.

There is therefore no need to wait for a change in the standard to return uncensored results.

## 8.3. Public perception

Although, to our knowledge, there is no study on public perception regarding the restitution of non-significant measurement results, several observations of principle can be made and a textbook case exists.

Studies have been done on scientists' understanding of the concept of statistically significant. They conclude that the overwhelming majority of researchers and engineers have an approximate or even completely erroneous conception of this idea (Belia et al., 2005; Greenland et al., 2016). There is little chance that the general public will correctly understand what the experts themselves struggle to explain.

# 9. CURRENT DISCUSSIONS ON STATISTICAL INFERENCE

"Statistically significant- don't use it, don't say it"- American statistician

## 9.1. An old discussion

Even if critics of the concept of "statistically significant" have raised objections from its inception, it has become almost universally accepted. Voices are regularly raised to recall or explain the faults and problems of this approach. (Giere, 1972; Harlow et al., 1997; Rozeboom, 1960). The first criticisms mainly concerned the difference in approach between Neyman and Pearson on the one hand and Fisher on the other. The two approaches are often grouped under the term Null Hypothesis Significance Testing (NHST) as we have mentioned. With the resurrection of the Bayesian paradigm, the philosophical and interpretative differences between it and frequentism have been the main front of criticism. Other criticisms have regularly been made of the totally arbitrary nature of the threshold used to declare a result as significant. The p-value is traditionally set at 0.05. There is no objective justification for this specific threshold (Cowles & Davis, 1982). The designer (Fisher) of the p-value never really gave an explanation on the origin of the threshold at 0.05. As we mentioned, this threshold is not the same in the field of radioactivity (0.025) or in particle physics. The interested reader can refer to Nickerson for an overview of the controversies before 2000 (Nickerson, 2000). Of course, the criticisms varied as the concept became more hegemonic in publications (for example, it established itself in the field of psychology until reaching a presence in 95% of publications). (Hubbard & Ryan, 2000))  More recently these criticisms of principle have intensified by focusing more on the consequences of the methodology.

## 9.2. The current « revolt »

In 2005, in a resounding publication (more than 8,300 citations), Ioannidis showed that a significant percentage of scientific publications are likely false. (Ioannidis, 2005). We will not explain the details of the reasoning here but you can find an educational explanation in the video. The problem is partly linked to statistical considerations and in particular to statistical "significance". As a result, this publication led to what we call the reproducibility crisis. Attempts to replicate results have been made, sometimes with astonishing results (out of 53 studies chosen as important on cancer, only 6 could be successfully reproduced). In 2006, the book "the cult of statistical significance" was published (Ziliak & McCloskey, 2008). Written by economists, it focuses its criticism on the fact that a statistically significant difference can correspond to a tiny effect, differentiating between what is statistically significant and what is practically significant. Under certain conditions, an effect can indeed be tiny but statistically significant and therefore of no real importance. It is clear that the term significant is not very happy for this reason. In 2016, the American Statistical Association took up the subject. This professional association has 18,000 members, historically the second oldest existing professional society in the United States. For the first time in its history, it published recommendations in the form of a declaration (Wasserstein & Lazar, 2016). We reproduce below the 6 recommendations:

1. ***P-values can indicate how incompatible the data are with a specified statistical model.***

2. *P-values do not measure the probability that the studied hypothesis is true, or the probability that the data were produced by random chance alone.*
3. *Scientific conclusions and business or policy decisions should not be based only on whether a p-value passes a specific threshold.*
4. *Proper inference requires full reporting and transparency.*
5. *A p-value, or statistical significance, does not measure the size of an effect or the importance of a result.*
6. *By itself, a p-value does not provide a good measure of evidence regarding a model or hypothesis.*

Initially, this crisis focused on scientific studies aimed at proving the existence of an effect or the effectiveness of treatment. She therefore focused on the concept of p-value. We saw previously that there is a close link between the p-value and the hypothesis testing approach in metrology. Through its use of the concept of "statistically significant", metrology is also fully concerned. If we translate these recommendations into the metrological context, the main points would be as follows::

- The significant nature of a measurement does not alone prove the presence of a signal.
- • Conversely, a non-significant result does not by itself prove the absence of a signal.
- No scientific conclusion should be based solely on the fact that a measurement exceeds the decision threshold.

Correct inference requires full restitution and total transparency of results.

The conclusion is immediate. In the context of metrology, a significant and non-significant result should not be treated differently! **The measurements must therefore always be reproduced in the same way, providing a result with its uncertainty.**

In a March 2019 issue of Nature, an article co-authored by more than 800 scientists from all disciplines (Amrhein et al., 2019) calls for the concept of "statistically significant" to be abandoned. The same month, the journal "American Statistician" published a special 400-page special issue with around fifty contributions (including big names in statistics) devoted solely to the subject of statistically significant. It is impossible to summarize all of these contributions and suggestions here, but the consensus seems to be on abandoning a differentiated treatment of significant and non-significant results. This arbitrary "dichotomization" is rejected in favor of a full restitution of the results and a mention of the p-value with possible comments.

This revolt, since this is how it has been described, has not yet reached the field of metrology but it seems inevitable to the author (and profitable for all) that this will come in the short or medium term.

# 10. APPLICATION EXAMPLES

## 10.1. Giant Clams measurements as part of surveillance in Polynesia

Nous allons reprendre ici, un exemple utilisé dans un précédent rapport. Pour plus de détail, il conviendra de s'y référer (Manificat, 2015).

Carried out since 1962 in Polynesia, radiological monitoring of the French environment, which currently concerns seven islands (Tahiti, Maupiti, Hao, Rangiroa, Hiva Oa, Mangareva and Tubuai) representative of the five archipelagos, consists of regularly taking samples of varied nature in the different environments (air, water, soil) with which the population may be in contact, as well as foodstuffs (Bouisset 2011, Bouisset 2014).

Regarding foodstuffs, the samples analyzed are representative of the food ration of Polynesians living in the five archipelagos of this territory, and come from the open sea marine environment, the lagoon marine environment and the terrestrial environment..

Among the marine samples, the giant clam (Tridacna maxima) is a filter-feeding organism collected and measured for many years as part of the IRSN monitoring program..

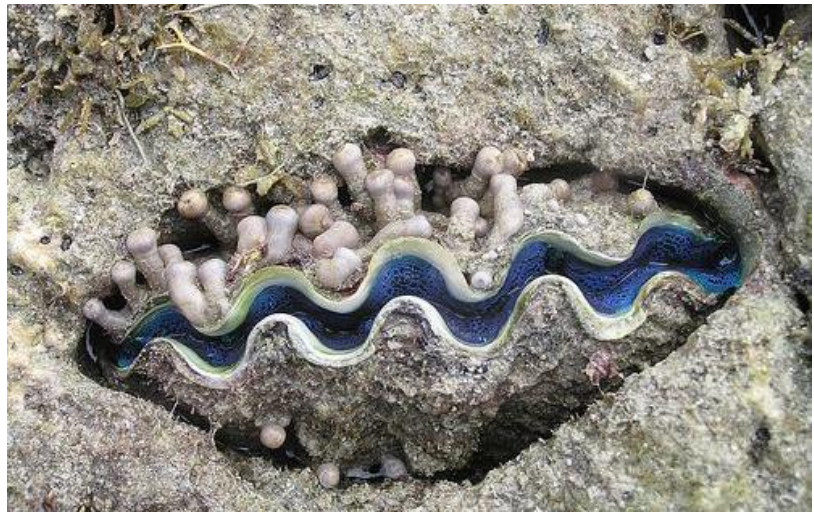

**Figure 10 - Picture of a giant clam in the Pacific**

Cesium 137 has obviously been fmonitored for many years.

Here is what the chronicle of the measurement results of cesium 137 in Bq/kg in the giant clams gives, adopting the standard restitution and therefore censoring the non-significant results..

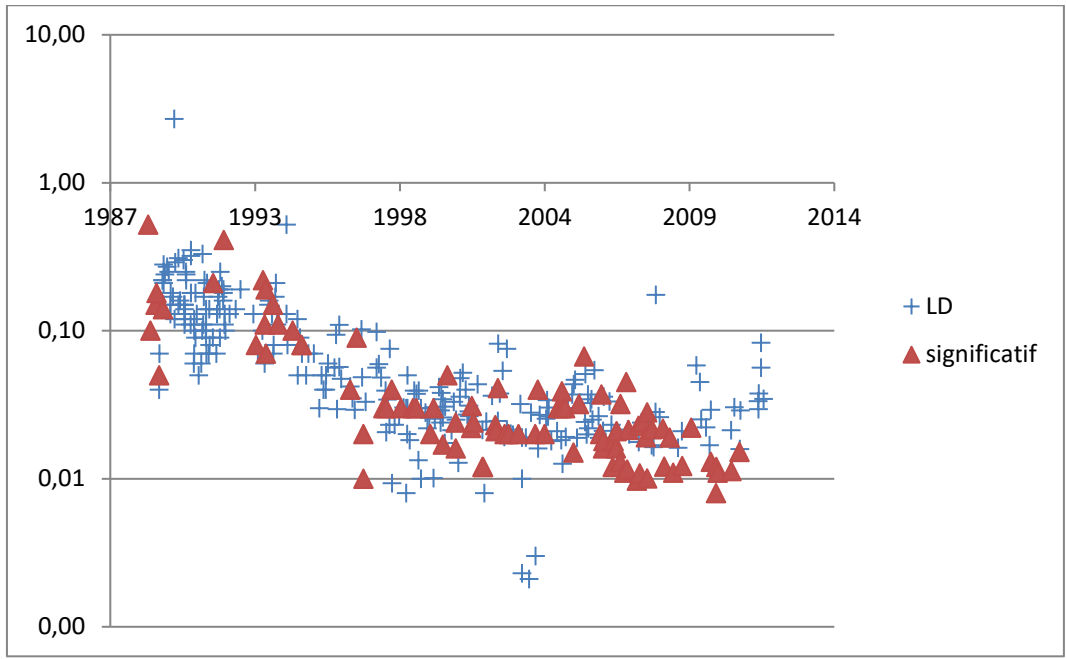

**Figure 11 - measurement results (in fresh Bq/kg) with censoring of clam samples between 1989 and 2012**

We can clearly discern a plateau in the results from the 2000s. However, the existence of this plateau does not correspond to a physical reality. There is no contribution of cesium 137 to Polynesia as shown by the water and air samples taken elsewhere. Cesium 137 must therefore decrease.

If we reanalyze the spectra to extract the non-significant results, we obtain the following curve:

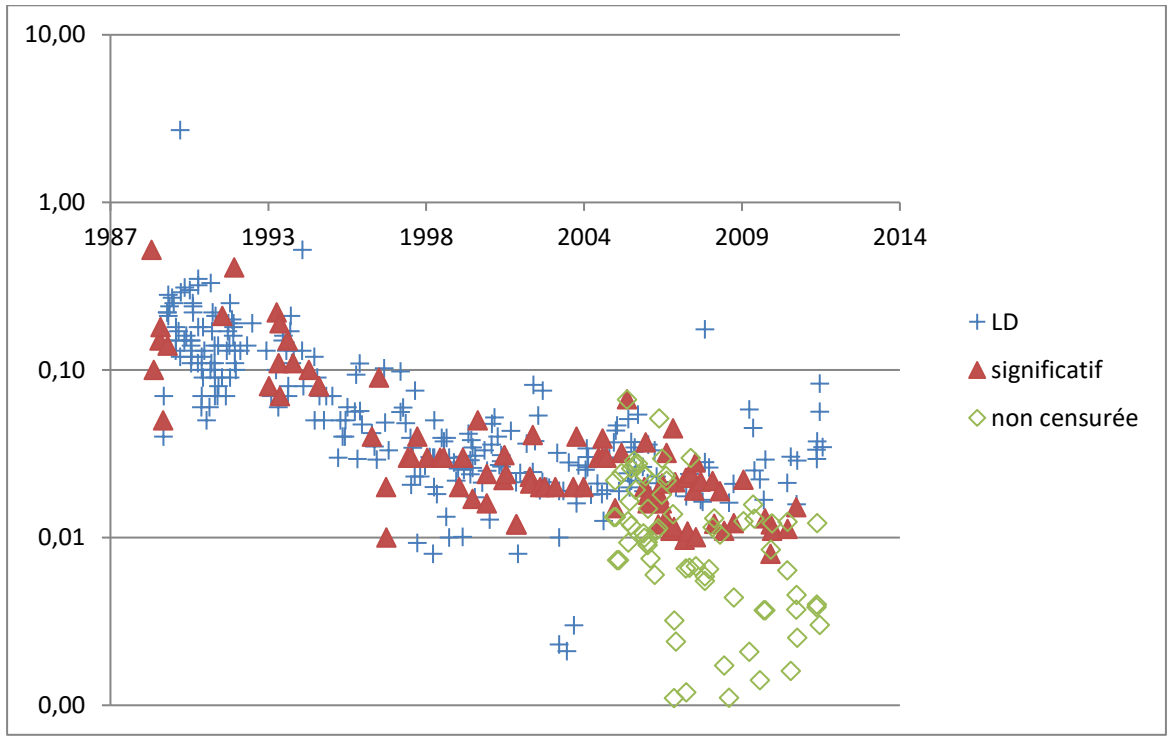

**Figure 12 - measurement results (in Bq/kg fresh) with censoring of giant clam samples between 1989 and 2004 and uncensored beyond**

We can clearly see the disappearance of this plateau and the continuation of the decline, which match physical reality. The uncensored results therefore make it possible to study long-term trends. On the other hand, the measurement of cesium 137 is done by gamma spectrometry which involves analyzing spectra. Even for these complex measurements, it is perfectly possible to extract uncensored results.

## 10.2. Tritium measurements in rainwater

Around forty rainwater samples were studied using three tritium measuring devices with detection capabilities ranging from the most efficient to the least efficient.

| Measurement devices | Number of measurements | Censorship rate | Non censored data provided | Comments |
|---|---|---|---|---|
| ALOKA | 46 | 4 % | no | These measurements will be considered as « true values » |
| TriCarb | 45 | 31 % | yes | |
| TriCarb « classical » | 38 | 85 % | yes | |

Aloka is a very efficient detector and the vast majority of tritium measurement results have been declared statistically significant according to the ISO standard. We can therefore consider these results to be a very good approximation of the true values. On the other hand, for the other two devices, the rates of censored values (not significant) are respectively 31% for tricarb and 85% for "classic" Tricarb. Work was done to extract the uncensored values from these two devices. A comparison is therefore possible between the "true" values of the Aloka and the censored and uncensored values of the two other measuring devices. The figure below summarizes the results.

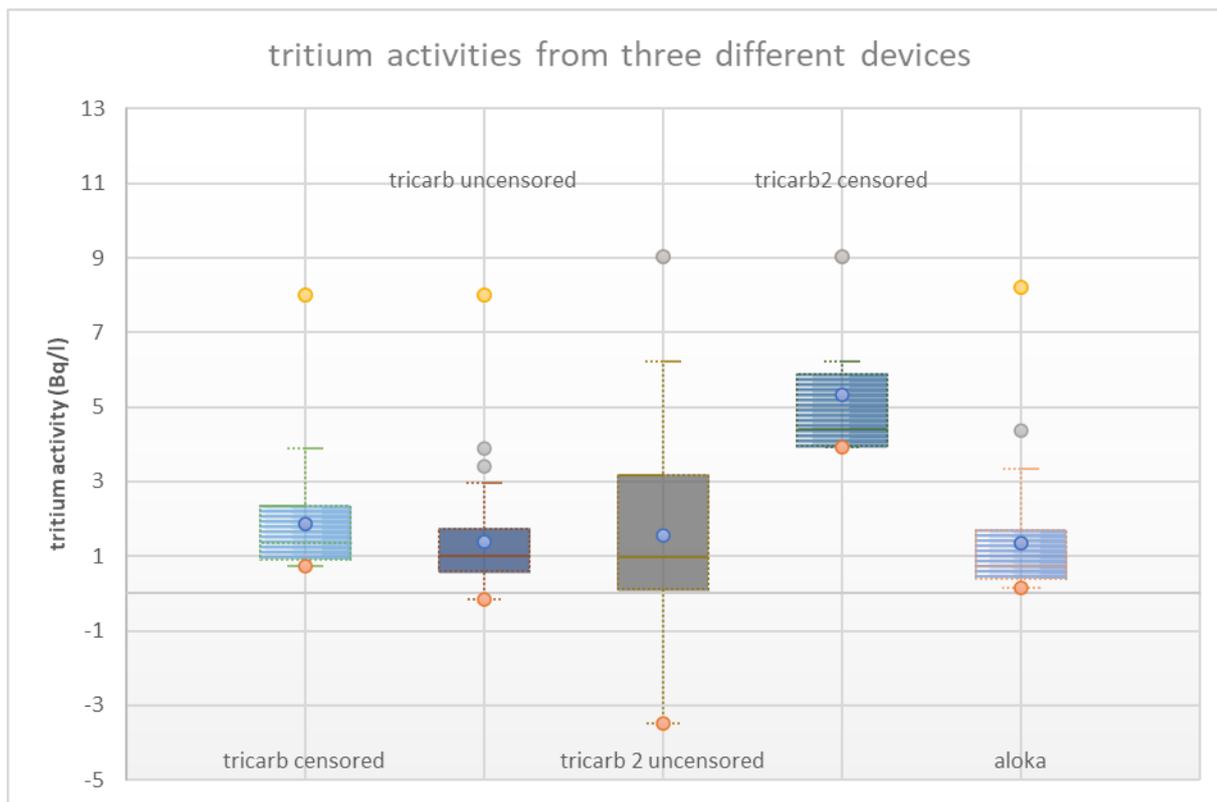

**Figure 13 - comparison of the results of censored and uncensored measurements of the three devices.**

We see that the censored values (first and fourth "box-plots") are biased compared to the true values. This is all the more true with a non performing detector where the censorship rate is close to 80%.

On the other hand, as soon as we integrate the non-significant values (second and third "box-plots") we find results comparable with the true values, provided that the negative values are included. These Tricarb values are logically more dispersed (greater uncertainty). Indeed, the performance of the detector is linked to the uncertainty in the reference, which will directly reflect when we substract from the values measured for the sample. It is possible to statistically compare the distributions of uncensored values of the three devices. Tests indicate that the three detectors give completely compatible values. A comparison with mathematical methods for exploiting censored results was also made. It unambiguously concludes that the uncensored results perform better.

This physical experiment ensures that the method presented in this report is not a mathematical artifice. We find the true values with consistent statistical performances.

## 10.3. Astrophysics

IIt should also be noted that the determination of these decision thresholds and detection limits corresponds exactly to that of the so-called « on-off » astrophysics problem. (or Li-ma problem) (Gillessen & Harney, 2005; Gregory, 2005; Li & Ma, 1983). This involves determining whether during an observation of an area of space, something was detected as significantly deviating from the background noise. The latter is generally determined by observing an area known to be free of the desired phenomenon. This problem is commonly encountered in gamma-ray burst research. The resolution method is generally to consider that the background noise parameter is known both from a frequentist and Bayesian point of view. This amounts to considering that in our notation $\mu = \lambda + \theta$, where $\lambda$ is known. To our knowledge, the method presented in this document is the first not to use this approximation. Let us also mention that the application of Bayesian methodology is done for this problem using a prior with positive support under the pretext of the non-physical nature of the negative parameters.

A detector points to a region (called « On ») suspected of containing a gamma ray source. For a time $t_{On}$, it is counting $N_{On}$ gamma photons. The detector is then pointed towards a region (called « Off ») where it is assumed that no gamma sources are present. during a time $t_{Off}$, $N_{Off}$ gamma photons are counted. Setting $\alpha = t_{On}/t_{Off}$, it is then possible to assume that the background noise in the « On » region possess the same distribution as in the « Off » region. This amounts to considering that in the « On » region the background noise is $\alpha N_{Off}$. We will assume that we are in a Poisson model, which is natural for rare events. It is then possible to say from equation (9), that the decision threshold $k$ verify:

$$\alpha_c = I_{1/2}(\alpha N_{off} + k + 1 - a, \alpha N_{off} + 1 - a)$$

In the field of astrophysics, it is customary to use the significance $S$ defined as the ratio between the signal and the standard deviation.

We therefore have, passing to the limit of large counts:

$$S = \frac{N_{On} - \alpha N_{Off}}{N_{On} + \alpha N_{Off}}$$

## 10.4. Particle physics

The problem of determining whether a signal deviates from the background noise for background distributions is also ubiquitous in particle physics. Events are counted in different configurations and the aim is to determine whether during experiments these counts deviate from the basic signal. This was particularly the case for the discovery of the Higgs boson. (van Dyk, 2014) where the equivalent of the probability $\alpha_c$ was chosen as worth approximately $10^{-6}$. The methodologies used are generally frequentist but also reject negative parameters considered as non-physical. Some researchers use Bayesian methods with of course strictly positive supported priors. (Gregory, 2005; James, 2006; Lista, 2016). Whether for Bayesian or frequentist methods, the background noise parameter is generally assumed to be known. For example, to deal with a counting problem (Lista, 2016), we assume that the background noise parameter is perfectly known $\lambda$, we are interested in the signal parameter $s$ and we have a distribution of the sample counts (signal + background noise):

$$p_G(n|s,\lambda) = \frac{(s+\lambda)^n}{n!} e^{-(s+y)}$$

This approach has been modified to take into account the fact that the background noise is not always perfectly known (Lista, 2016).

In the work presented here, the background noise parameter is not assumed to be known and only observations are taken into account beyond the mathematical model used (likelihood function). If we assume that the background noise is Gaussian with a mean $y$ and a variance $2\sigma_B^2$, we therefore show using the results of this document that the level of "significance" will in fact be:

$$Z = \frac{n-y}{\sqrt{2\sigma_B^2 + n - y}}$$

The significance level is defined as the equivalent number of standard deviations for the measurement ((Lista, 2016). A Confidence index of à 5% will correspond to Z=1,6, 2,5% for Z~2 and it is customary to declare a discovery in particle physics tor Z=5 ($\alpha_c$=3,7. $10^{-5}$%).

Another point of view would be to consider the Skellam distribution in its Gaussian limit. If we have a measurement $y$ for the reference and $x$ for the sample, the level of significance will then be:

$$Z = \frac{x-y}{\sqrt{x+y}}$$

This is perfectly consistent with the fact that the difference between a Poisson distribution of parameter $\mu$ and a Poisson distribution of parameter $\lambda$ is a Skellam distribution with mean $\mu - \lambda$ and variance $\lambda + \mu$ (Skellam, 1946). Using maximum likelihood estimators, we obtain the previous result.

It is also possible from the results of 7.2.1, 7.3.2 to determine the confidence interval for the parameter $\theta$. On the other hand, it is not possible to guarantee both the coverage probability and the positivity of a sample position parameter. We can give a confidence interval for a parameter $\theta$ which will have the right coverage properties but it will potentially include negative values because it is the parameter of the net random variable. The alternative is to give an interval containing only positive values but it will then not have the good coverage properties.

# 11. CONCLUSIONS

The low-level measurement problem can be approached in at least two different ways. Classical statistics allows the use of the Neyman-Pearson lemma. This allows us to obtain the optimal hypothesis test. The other approach is that of Bayesian statistics. With very few exceptions, the application of Bayesian methodology to metrology issues is inevitably done with strictly positive support priors for measurands which are ideally positive. ((Bergamaschi et al., 2013; Bochud et al., 2007; IAEA, 2017; ISO, 2010a; Korun et al., 2014, 2016; Laedermann et al., 2005; Lira, 2009; Little, 1982; Michel, 2016; Miller et al., 2002; Rivals et al., 2012; A. Vivier et al., 2009; Weise et al., 2006; Zähringer & Kirchner, 2008; Zorko et al., 2016)). It seemed natural to limit ourselves to a domain where the desired parameter is positive. As we have seen, this confuses the true signal with the net signal and biases the results with inadequate statistical performance. Using a simple criterion very similar to that of classical statistics, we were able to formulate a way to determine the significant nature of a measurement. This direct use of confidence or credibility intervals makes it possible to obtain an optimal test according to the Neyman-Pearson methodology and largely reconciles classical and Bayesian approaches. Numerical simulations confirm the statistical performance of these criteria.

The recommendations and conclusions deduced from this document are therefore as follows:

- • Measurements below the decision threshold should not be censored.
- The difficulties encountered fall within the framework of the difficult problem of nuisance parameters (how to find θ without knowing λ). We propose a conditional approach for the frequentist approach and a marginal approach for the Bayesian approach. The two approaches converge for sufficient count values. This method works with members of the family of natural exponentials with quadratic variance (NEF-QVF) which bring together a large part of the probability densities commonly used in metrology. The use of a prior with positive support such as the Heavyside prior should be avoided in the case where we admit the possibility of fluctuating measurements around the reference (negative measurements).
- • Rejecting the negative part of the confidence interval like [-a,b] under the pretext of its non-physical character is nonsense. The parameter accessible to measurement by inference is not the "true" signal but the "net" signal which differs from it by its greater dispersion. This "net" signal has no physical reason to be positive and under no circumstances should the intervals be truncated by excluding the negative part.
- Obtaining the probability density of the parameter θ based on observations $z$, $p(\theta|z)$ is sufficient to define an estimation interval for the parameter and determine the significant nature or not of the measurement. This also provides an upper bound for the parameter which will depend on the measurement and will be more precise than the detection limit. The hypothesis tests which make it possible to define the characteristic limits of the model (decision thresholds, detection limits) are intrinsic to the construction of the estimation intervals. Mere knowledge of this interval is sufficient. This also makes it possible to avoid several paradoxes (inconsistency between the estimation intervals and the hypothesis tests as mentioned in paragraph 7.9, divergence of the detection limit as mentioned in 7.3.6). It is therefore necessary and sufficient to provide a result with its uncertainty. This is in line with current concerns in the field of statistics on the use of the significant nature or not of a result. No other test can have better statistical performance.

- For Gaussian distributions and more generally for position parameters, the equality between the confidence intervals and the credibility intervals ensures that these intervals and the associated characteristic limits have the right statistical properties (coverage probability). For the case of radioactivity, numerical experiments show that the decision threshold determined from the estimation intervals is better than that of Currie and ISO11929. In fact, Neyman Pearson's lemma guarantees that this is the case.
- The confidence intervals contain all the necessary information and there is no need to censor the data• The ISO 11929 standard is inadequate and its theoretical foundation is erroneous through the use of a Heavyside prior and the use of frequentist characteristic limits on Bayesian distributions. It leads to underestimated decision thresholds.
- • The method proposed here gives substantially the same results for a frequentist or Bayesian approach. The (small) differences at low count values can be explained by the choice of the prior. These differences reflect the epistemic uncertainty of the lack of measurement data.
- • The method can easily be used to determine characteristic limits on other metrological techniques such as biological dosimetry, electron paramagnetic resonance, particle physics or astrophysics.

Many avenues of work exist. Applying these methodologies to cases where both location and dispersion are unknown would lead to Student distributions and it would be useful to calculate their characteristic limits. The extension of this methodology to techniques requiring calibration curves is another avenue. Furthermore, certain paradoxes identified in the use of Bayesian methods in metrology (Attivissimo et al., 2012) would perhaps see avenues of resolution open up.

## 12. ACKNOWLEDGEMENTS

I thank Alain Vivier for introducing me to this problem and for the fierce discussions which forced me to clarify my ideas. I also thank the strange fate that gave me the opportunity and time to think about this question. Finally I thank a disciple of Basu who will recognize himself.

# ANNEXE 1     PRODUCT AND CONVOLUTION OF GAUSSIAN DISTRIBUTIONS

## 1.1 Product

If f and g are both gaussians distributions :

$$f(x) = \frac{1}{\sqrt{2\pi}\sigma_f} e^{-\frac{(x-\mu_f)^2}{2\sigma_f^2}} \text{ et } g(x) = \frac{1}{\sqrt{2\pi}\sigma_g} e^{-\frac{(x-\mu_g)^2}{2\sigma_g^2}}$$

with $\sigma_{fg} = \sqrt{\frac{\sigma_f^2 \sigma_g^2}{\sigma_f^2+\sigma_g^2}}$ and $\mu_{fg} = \frac{\mu_f \sigma_f^2 + \mu_g \sigma_g^2}{\sigma_f^2+\sigma_g^2}$

the product will be :

$$f(x)g(x) = \frac{1}{\sqrt{2\pi}\sigma_{fg}} e^{\left[-\frac{(x-\mu_{fg})^2}{2\sigma_{fg}^2}\right]} \frac{1}{\sqrt{2\pi(\sigma_f^2+\sigma_g^2)}} e^{\left[-\frac{(\mu_f-\mu_g)^2}{2(\sigma_f^2+\sigma_g^2)}\right]}$$

## 1.2 Convolution

If f and g are both gaussians distributions :

$$f(x) = \frac{1}{\sqrt{2\pi}\sigma_f} e^{-\frac{(x-\mu_f)^2}{2\sigma_f^2}} \text{ et } g(x) = \frac{1}{\sqrt{2\pi}\sigma_g} e^{-\frac{(x-\mu_g)^2}{2\sigma_g^2}}$$

Then their convolution will be (Bromiley, 2003) :

$$f \oplus g(x) = \int f(t-x)g(t)dx = \frac{1}{\sqrt{2\pi(\sigma_f^2+\sigma_g^2)}} e^{\left[-\frac{(x-(\mu_f+\mu_g))^2}{2(\sigma_f^2+\sigma_g^2)}\right]}$$

# ANNEXE 2 TEST REPORTS

SL 31  **3H Combustion Environnement**  Monophase : 15,00 ml
P# 3HCombEnv  Prise d'essai : 1,00 ml
**1+MPLIQ**  U(Pe) : 0,04 ml (k=2)
Ech. N° 1  Fact. correctif : 1,00
Essai N° 1  *Recomptage*  Rdt comb : 100,00 %
3.8.b  U(Rdt comb) : 7,00 % (k=2)

| LUM | $N_{brut}$ (cpm) 0 à 8 KeV | | $N_{brut}$ (cpm) 15 à 80 keV | | $N_{brut}$ (cpm) 80 à 2000 keV | | Temps (min) | Tsie | Date de comptage |
|---|---|---|---|---|---|---|---|---|---|
| 4 | 1,960 | bon | 3,020 | bon | 4,040 | bon | 50 | 521,17 | 24/07/2019 13:12:09 |
| 4 | 2,540 | bon | 3,200 | bon | 4,280 | bon | 50 | 522,39 | 25/07/2019 09:26:28 |
| 4 | 1,980 | bon | 3,160 | bon | 3,420 | douteux | 50 | 519,07 | 26/07/2019 05:40:45 |
| 4 | 2,240 | bon | 3,380 | bon | 4,640 | bon | 50 | 521,15 | 27/07/2019 01:55:05 |
| | | | | | | | | | |
| | | | | | | | | | |
| | | | | | | | | | |
| | | | | | | | | | |
| **Nbrut** | 2,180 | cpm | 3,190 | cpm | 4,095 | cpm | 200 | 520,95 | 25/07/2019 19:33:37 |
| MP | 2,172 | cpm | 3,008 | cpm | 4,305 | cpm | | | |
| **Nnet** | 0,008 | cpm | 0,183 | cpm | -0,210 | cpm | | | |
| Sigma | 0,289 | cpm | 0,340 | cpm | 0,407 | cpm | | | |

SL 31  **3H Combustion Environnement**  Monophase : 15,00 ml
P# 3HCombEnv  Prise d'essai : 2,00 ml
**15+S19EEA21-98B1**  U(Pe) : 0,08 ml (k=2)
Ech. N° 15  Fact. correctif : 1,00
Essai N° 1  *Recomptage alcoo*  Rdt comb : 100,00 %
3.8.b  U(Rdt comb) : 7,00 % (k=2)

| LUM | $N_{brut}$ (cpm) 0 à 8 KeV | | $N_{brut}$ (cpm) 15 à 80 keV | | $N_{brut}$ (cpm) 80 à 2000 keV | | Temps (min) | Tsie | Date de comptage |
|---|---|---|---|---|---|---|---|---|---|
| 4 | 2,140 | bon | 2,720 | douteux | 3,860 | bon | 50 | 450,27 | 25/07/2019 03:17:28 |
| 3 | 2,540 | bon | 3,340 | bon | 4,280 | bon | 50 | 444,74 | 25/07/2019 23:32:06 |
| 3 | 2,760 | bon | 3,480 | bon | 4,200 | bon | 50 | 450,40 | 26/07/2019 19:46:30 |
| 3 | 2,480 | bon | 3,600 | bon | 4,160 | bon | 50 | 454,92 | 27/07/2019 16:00:45 |
| | | | | | | | | | |
| | | | | | | | | | |
| | | | | | | | | | |
| | | | | | | | | | |
| **Nbrut** | 2,480 | cpm | 3,285 | cpm | 4,125 | cpm | 200 | 450,08 | 26/07/2019 09:39:12 |
| MP | 2,180 | cpm | 3,190 | cpm | 4,095 | cpm | | | |
| **Nnet** | 0,300 | cpm | 0,095 | cpm | 0,030 | cpm | 46,27 | 841,00 | Tsie de 0 à 8 KeV |
| SD | 0,289 | cpm | 0,350 | cpm | 0,397 | cpm | | | |
| LD | 0,599 | cpm | 0,000 | cpm | | | | | |

**CALCUL FENÊTRE : 0 à 8 KeV**  Radionucléide : 3H  *Tsie OK*
Période(j) : 4,50E+03

Rdt (3H) = 42,94 ± 3,90 % (k=2) le 25/08/2017

Nnet (3H) = 0,300 ± 0,305 cpm (k=2)

Activité brute = 0,012 Bq/pe  U(A)/A = 102,48 % (k=2)
SD = 5,616 Bq/l  A = 0,012 Bq/pe
LD = 11,622 Bq/l  **A** = **5,822E+00 Bq/l**

## ANNEXE 3    HYPERGEOMETRIC FUNCTION

We saw that we could express $p_\Theta(\theta|x,y) = \frac{1}{\Gamma(x+1)\Gamma(y+1)} \iint \lambda^y e^{-\lambda} \mu^x e^{-\mu} \delta(\theta + \mu - \lambda) \, d\mu d\lambda$

It is then necessary to differentiate the case $\theta > 0$ from $\theta < 0$ (Papoulis, 2002).

$p_\Theta(\theta|x,y) = \frac{1}{\Gamma(x+1)\Gamma(y+1)} \int_0^\infty (\mu + \theta)^x e^{-(2\mu+\theta)} \mu^y \, d\mu$ pour $\theta > 0$

$p_\Theta(\theta|g,b) = \frac{1}{\Gamma(x+1)\Gamma(y+1)} \int_0^\infty \lambda^x e^{-(2\lambda-\theta)} (\lambda - \theta)^y \, d\lambda$ pour $\theta < 0$

The above integrals can be expressed in the form of a confluent hypergeometric Tricomi function (Bergamaschi et al., 2013):

$$U(a,b,z) = \frac{1}{\Gamma(a)} \int_0^\infty e^{-zt} t^{a-1} (1+t)^{b-a-1} \, dt$$

For $\theta > 0$

$$p_\Theta(\theta|x,y) = \frac{1}{\Gamma(x+1)\Gamma(y+1)} e^{-\theta} \int_0^\infty (\mu+\theta)^x e^{-2\mu} \mu^y \, d\mu$$

$$= \frac{1}{\Gamma(x+1)\Gamma(y+1)} e^{-\theta} \theta^{x+y} \int_0^\infty \left(\frac{\mu}{\theta}+1\right)^x e^{-2\theta \frac{\mu}{\theta}} \left(\frac{\mu}{\theta}\right)^y \theta \, d\left(\frac{\mu}{\theta}\right)$$

$$= \frac{1}{\Gamma(x+1)\Gamma(y+1)} e^{-\theta} \theta^{x+y+1} \int_0^\infty (t+1)^x e^{-2\theta \cdot t} (t)^y \, dt$$

$$= \frac{1}{\Gamma(x+1)} e^{-\theta} \theta^{x+y+1} U(y, x+y+2, 2\theta)$$

In the same way, for $\theta < 0$ :

$$p_\Theta(\theta|x,y) = \frac{1}{\Gamma(x+1)} e^{\theta} \theta^{x+y+1} U(x, x+y+2, -2\theta)$$

Tables or mathematical software then make it possible to calculate this expression for all the values of $\theta$, $x$ and $y$.

It is possible to calculate the decision threshold from this expression:

$$p_\Theta(\theta > 0|x,y) = 1 - p_\Theta(\theta < 0|x,y) = 1 - \alpha_c = \frac{1}{\Gamma(y+1)} \int_0^\infty e^{-\theta} \theta^{x+y+1} U(y, x+y+2, 2\theta) d\theta$$

Mathematics tables (Gradshteyn et al., 2000) give us :

$$\int_0^\infty e^{-s\theta}\theta^{b-1}U(a,c,\theta)d\theta = \frac{\Gamma(b)\Gamma(b-c+1)}{\Gamma(a+b-c+1)}F(b,b-c+1;a+b-c+1;1-s)$$

Where $F(a,b;c;x)$ is the hypergeometric function.

By changing the variable $v = 2\theta$, we get:

$$\alpha_c = \frac{1}{\Gamma(x+1)}\int_0^\infty e^{-v/2}(v/2)^{x+y+1}U(y,x+y+2,v)dv$$

Using the previously mentioned formula, we obtain:

$$\alpha_c = \frac{1}{2^{x+y+1}}\frac{1}{\Gamma(x+1)}\frac{\Gamma(x+y+2)}{\Gamma(y+1)}F(x+y+2,1;y+1;1/2)$$

If we know $y$, this amounts to finding k such that:

$$\alpha_c = \frac{1}{2^{2y+k+1}}\frac{1}{\Gamma(y+k+1)}\frac{\Gamma(2y+k+2)}{\Gamma(y+1)}F(2y+k+2,1;y+1;1/2)$$

The advantage of this type of formula is that it does not require any knowledge other than y to obtain an expression for k.

# ANNEXE 4    NEYMAN PEARSON LEMMA

Consider a null hypothesis $H_0$. This is an assertion about a statistical distribution that we wish to test, generally in the form of the absence of an effect.

We then consider the alternative hypothesis $H_a$. Our goal is then to determine, using a test on observations, which hypothesis is most compatible with the data. Let us insist on the fact that Neyman Peason's approach, which the metrological criteria are based on, necessarily requires the definition of this alternative hypothesis.

In the null hypothesis, we assume that it is the noise($B'$) which will generate the measurement $x$ (we therefore have in this hypothesis θ=0). We know that if we obtained $y$ lduring a first measurement then a second measurement is distributed according to $p_N$ :

$$p_{G|B}^{H_0}(x|y) = p_N(x|y) = \frac{e^{-\frac{(x-y)^2}{4\sigma_B^2}}}{\sqrt{4\pi}\sigma_B} = \frac{e^{-\frac{z^2}{4\sigma_B^2}}}{\sqrt{4\pi}\sigma_B} = p_N(z|0, 2\sigma_B^2)$$

For the alternative hypothesis($\theta > 0$),

$$p_{G|B}^{H_a}(x|y) = p_N(z|\theta, 2\sigma_B^2) = \int_{-\infty}^{\infty} p_S(z-w|\theta) p_N(w|\varepsilon, 2\sigma_B^2) dw = \int_{-\infty}^{\infty} p_S(x-y'|\theta) p_N(y'|y, 2\sigma_B^2) dw$$

The Neyman-Pearson approach will seek to determine whether it is likely that θ=0 through the use of likelihood ratios (Lehmann & Romano, 2005a) :

$$\Lambda(z|\theta) = \frac{p_{H_0}(z)}{p_{H_a}(z|\theta)} = \frac{\frac{e^{-\frac{z^2}{4\sigma^2}}}{\sqrt{4\pi}\sigma}}{\frac{e^{-\frac{(z-\theta)^2}{4\sigma^2}}}{\sqrt{4\pi}\sigma}} = e^{-\frac{(2z\theta-\theta^2)}{4\sigma^2}}$$

*(30)*

Neyman Pearson's lemma proves that the most powerful test is the one that rejects $H_0$ for the benefit of $H_a$ when $\Lambda(z|\theta) \leq c_\alpha$, with $c_\alpha$ such that:

$$p_{H_0}(\Lambda(z|\theta) \leq c_\alpha) = \alpha$$

This means that the set of z such that $\Lambda(z|\theta) \leq c_\alpha$ is equivalent to the set of z such that $z > z_c$, with $z_c$ determined by :

$$p_{G-B}^{H_0}(z > z_c) = \alpha$$

By the most powerful test, we mean the test that will minimize the false negative rate for a fixed false positive rate. Note that the ratio of the equation (30) is decreasing as z increases, if θ is positive. This condition is met in the present case since we are interested in a positive physical quantity. We can even add that this ratio decreases monotonically. The Karlin-Rubin theorem (Karlin & Rubin, 1956) which is an extension of the Neyman-Pearson lemma, then assures us that it is the uniformly most powerful test. That is to say, if we want to test whether $\theta > 0$, the criterion $z > z_c$ with $z_c$ such that

$$\alpha_c = \int_{z_c}^{\infty} p_{H_0}(z)dz = \int_{z_c}^{\infty} p_N(z|0)dz$$

Will lead to the test minimizing the false negative rate, while fixing the false positive rate. Note that we are in fact testing whether $\theta > 0$ which corresponds to testing wether $\mu > \lambda$.

The proportion of false negatives for a given value of the parameter θ will then be:

$$\beta_c = \int_0^{z_c} p_N(z|\theta)dz$$

No other statistical test of the same level $\alpha_c$ will be able to give a smaller $\beta_c$ (the power of the test).

Note that such a definition is perfectly compatible with that of the ISO (ISO, 2010a) which does not specify the type of hypothesis test to be carried out.

Under the null hypothesis, the proportion of measures exceeding this desired level $z_c$ must therefore reach a value α that we set a priori. If the measurement exceeds this threshold thus determined, we can reject this hypothesis. Only $100\alpha\%$ The measures would statistically exceed this level $z_c$ if the parameter was null. This amounts to having $100\alpha\%$ probability of false positive if the parameter was zero.

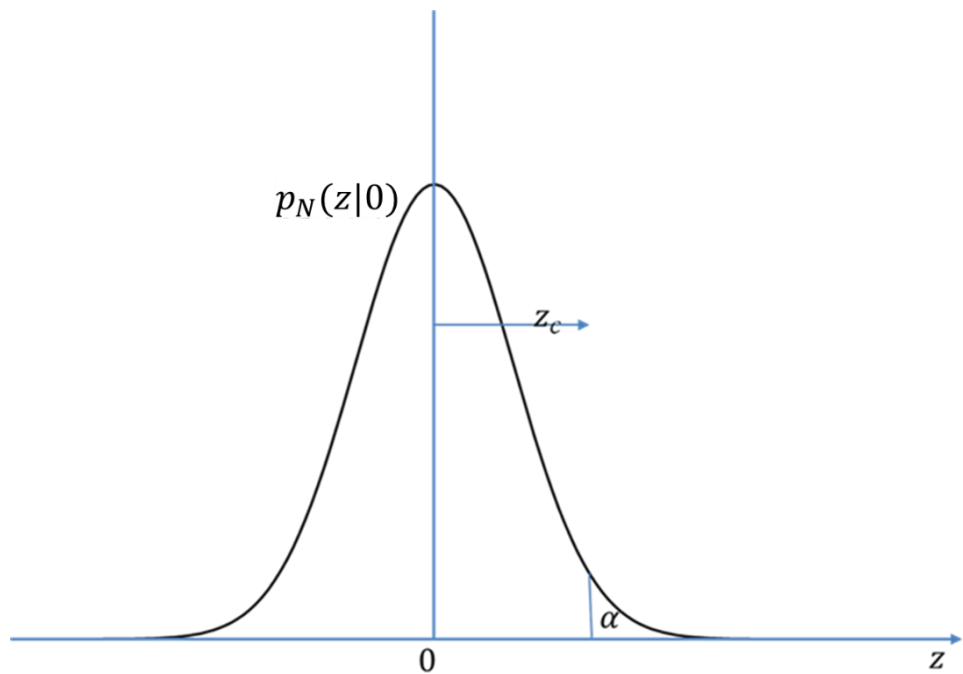

**Figure 14 - Schematic diagram of the frequentist determination of the decision threshold**